\documentclass[pre,a4paper,superscriptaddress,twocolumn,showpacs,amsmath,amssymb,floatfix,amstex]{revtex4}%

\newcommand{\fig}[1]{Fig.~\ref{#1}}         						%
\usepackage{graphicx,graphics,rotating}	% Include figure files
\usepackage{dcolumn}
\usepackage[hidelinks]{hyperref}
\usepackage{mathptmx}
\usepackage{array}
\usepackage{verbatim}
\usepackage{amsmath}
\usepackage{mathrsfs}
\usepackage{xcolor}
\usepackage{bbold}

\newcommand{\ket}[1]{| #1\rangle}                       %
\newcommand{\bra}[1]{\langle #1|}                        %

\newcommand{\lam}[0]{|\lambda_j|}

\newcommand{\Pl}[0]{\Psi}  
\newcommand{\Pz}[0]{\Psi^{(0)}}  
\newcommand{\rrp}{R_+^m}
\newcommand{\rrm}{R_-^m}

\newcommand{\lev}{\Psi_n^-}
\newcommand{\calH}{\mathcal{H}}
\newcommand{\tU}{\widetilde{U}}
\newcommand{\hole}{\Omega}
\newcommand{\proj}{1-\Pi_0}

\definecolor{mygreen}{rgb}{ .01 , .52 ,  .19}
\definecolor{myred}{rgb}{ .71 , 0 ,  0}
\definecolor{oli}{rgb}{ .99 , .39 ,  .19}
\definecolor{mygray}{rgb}{ .40 , .40 ,  .40}
\definecolor{purple}{rgb}{ .70 , .20 ,  .50}
\definecolor{lblue}{rgb}{ .30 , .50 ,  .90}

\usepackage{soul}
\soulregister\ref7
\begin{document}
\title{Multifractality of open quantum systems}
\author{Agust\'{\i}n M. Bilen}
\affiliation{Instituto de Investigaciones Físicas de Mar del Plata (IFIMAR), Facultad de Ciencias Exactas y Naturales, Universidad Nacional de Mar del Plata, CONICET, 7600 Mar del Plata, Argentina}
\author{Ignacio Garc\'{\i}a-Mata}
\affiliation{Instituto de Investigaciones Físicas de Mar del Plata (IFIMAR), Facultad de Ciencias Exactas y Naturales, Universidad Nacional de Mar del Plata, CONICET, 7600 Mar del Plata, Argentina}
\author{Bertrand Georgeot}
\affiliation{%
Laboratoire de Physique Th\'eorique, IRSAMC, Universit\'e de Toulouse,
CNRS, UPS, France}
\author{Olivier Giraud}
\affiliation{LPTMS, CNRS, Univ. Paris-Sud, Université
Paris-Saclay - 91405
Orsay, France}
\date{\today}% 
%%%
\begin{abstract}
We study the eigenstates of open maps whose classical dynamics is 
pseudointegrable and for which the corresponding closed quantum system 
has multifractal properties. Adapting the existing general framework 
developed for open chaotic quantum maps, we specify the relationship between the 
eigenstates and the classical structures, and we quantify their 
multifractality at different scales. Based on this study, we conjecture 
that quantum states in such systems are distributed according to a hierarchy of classical structures, but these states are multifractal instead of ergodic at each level of the hierarchy. This is visible for sufficiently 
long-lived resonance states at scales smaller than the classical 
structures. Our results can guide experimentalists in order to observe 
multifractal behavior in open systems.
\end{abstract}
\pacs{05.45.Df, 05.45.Mt, 71.30.+h, 05.40.-a}
%03.67.Ac 	Quantum algorithms, protocols, and simulations
%05.40.-a       Fluctuation phenomena, random processes, noise, and Brownian motion
%05.45.Df 	Fractals 
%05.45.Mt 	Quantum chaos; semiclassical methods
%05.40.-a       Random processes
%64.60.al       Fractals in phase transitions
%71.30.+h 	Metal-insulator transitions and other electronic transitions
%
%03.65.Yz	Decoherence; open systems; quantum statistical methods
%--
%03.67.-a	Quantum information (see also 42.50.Dv Quantum state 
% engineering and measurements; 42.50.Ex Optical implementations 
% of quantum information processing and transfer in quantum optics)
%--
%05.45.Mt	Quantum chaos; semiclassical methods
\maketitle

%************************************************************************************************************************
\section{Introduction}
%************************************************************************************************************************
Multifractals are a generalization of the simpler, better known fractals 
\cite{mandelbrot1982,falconer}.
Generically, a physical object or observable is multifractal when a 
single fractal dimension is not enough to describe the patterns that 
repeat themselves at every scale. Originally introduced to treat 
dissipation fluctuations in fluid turbulence \cite{mandelbrot1974}, it 
has proven a valuable tool to characterize many classical complex 
phenomena and has recently found application in the quantum realm as 
well. Notably, at the critical point of certain quantum phase 
transitions, like the metal-insulator transition of the celebrated 
Anderson model \cite{anderson1958absence}, wave functions exhibit 
nontrivial fluctuations compatible with a multifractal description 
\cite{evers2008anderson}. Other noteworthy systems in which 
multifractality is present are pseudointegrable systems \cite{giraud2004,bogomolny,MGG,MGGG,MGGG12,GarciaMata:2009fk}, ground states of spin chains \cite{Atas2012}, 
systems with a many-body localization transition \cite{monthus2016many,Serbyn2017,mace2018}, 
and the 
related problem of random graphs with disorder \cite{biroli2012difference,kravtsov2015random,tikhonov2016fractality,tikhonov2016anderson,altshuler2016nonergodic,ourPRL2017,kravtsov2018non,biroli2018delocalization,tikhonov2019critical}.%\cite{graphs}.
 While much progress has been made on the theoretical side 
\cite{mirlin2000,evers2008anderson,rodriguez2009multifractal,rodriguez2011multifractal,mirlin1996prbm,mendezbermudez2012prbm,Fyodorov_2009,indians1,indians,garciagarcia,Bogomolny1999Models,giraud2004,bogomolny,MGG,MGGG,BogGirPRL,BogGir11,BogGir12,MGGG12}, 
the possibility of observation in experiments has remained difficult (with, however, some indirect characterizations \cite{richardella2010,lemarie2010,lopez2013,sagi2012}), 
mainly due to the fact that in a real experimental setting there are 
perturbations and imperfections that cannot be completely controlled.
One first attempt to solve this problem %narrow this gap
consisted in studying the extent to which the multifractal character of 
a quantum state persists when the original system is subjected to 
different types of modifications. This direction was pursued in detail 
in \cite{RemyPRL2014,RemyPRE2015}, where
it was shown that there are two main scenarios: one where there is a 
characteristic scale beyond which multifractality survives, and another in which multifractality is destroyed at all scales.

A related question to be considered is that measurement schemes often 
involve opening the system in some way. It then becomes important to determine how much of the original structure of the closed system remains 
when it is opened.
%%%
First steps in this direction were taken in 
\cite{mendez2005wigner,MGGG12,MendezBermudez2014} by studying 
the Wigner delay times and the multifractality of wave packets.

Here we will consider quantum maps that
are the quantized counterparts of classical area-preserving maps 
\cite{berry1979quantum}.
They have been extensively exploited, due to their inherent simplicity, 
to study generic quantum properties, particularly of systems with chaotic classical dynamics. \emph{Open} quantum maps, in which the evolution is no longer unitary and resonance 
eigenstates have a finite lifetime, are a natural extension  to be  considered in a 
realistic set up. There has been much interest and 
progress in determining the spectral as well as the eigenstate 
properties of these systems in the chaotic case 
\cite{casati,lu,schomerus,Nonnenmacher2005,schomerus2005,keating,Nonnenmacher_2007,dima2008,novaes2009,Kopp2010,ermann2012,altmann2013,Korber2013,altmann2015,ketzmerick} 
(see \cite{novaes} for a more extensive list of references). As regards the 
distribution of resonance eigenstates, it was found in \cite{keating} 
that the weight of these on certain classical sets determined by the 
classical escape dynamics is directly related in the semiclassical limit 
to the norm of the corresponding eigenvalues. Also, in \cite{ketzmerick} 
a relevant classical hierarchy of structures was identified and a 
family of {conditionally} invariant measures was explicitly constructed to explain the 
eigenstate distribution.

In this work we  consider a family of quantum maps whose classical dynamics is pseudointegrable \cite{richens1981}. This is reflected in the spectrum, with 
intermediate level-spacing statistics, but also in the eigenfunctions 
that are delocalized but multifractal 
\cite{giraud2004,bogomolny,MGG,GarciaMata:2009fk,MGGG,MGGG12}.
Our objective is to describe the structure of eigenfunctions and reveal 
how multifractality is affected when the map is opened. {In particular, it is to understand whether multifractality survives at some scales or is destroyed altogether, in line with the possible scenarios mentioned above \cite{RemyPRL2014,RemyPRE2015}. Our main findings are the following: we generalize the theory of 
\cite{casati,lu,schomerus,Nonnenmacher2005,schomerus2005,keating,Nonnenmacher_2007,dima2008,novaes2009,Kopp2010,ermann2012,altmann2013,Korber2013,altmann2015,ketzmerick}, 
which was built for chaotic systems, to this new type of dynamics. As in the chaotic case, we show that the distribution of eigenstate components is conditioned by a classical hierarchy of phase space structures. However, in the chaotic case the wave functions were ergodic on each level of the hierarchy, whereas in our case we formulate the conjecture that wavefunctions are multifractal on these structures. We support 
our conjecture using extensive numerical simulations, with some caveats: very short-lived resonances are too close to the degenerate subspace to be reliably analyzed in this respect; additionally, the phase space structures considered should be large enough compared to $\hbar$ in the relevant phase space direction for multifractality to be seen. Moreover, since our multifractal analysis requires us to work at a fixed system size (and hence fixed effective Planck's constant), we are led to consider a way of modifying the previously developed theory in cases in which the characteristic size of the classical regions is comparable to $\hbar$. We suggest that the resulting adaptation could be valuable in general in situations in which one is forced to cope with finite-$\hbar$ effects.}

%{Using 
%multifractal analysis along with results from the existing semiclassical theory for resonance eigenstates 
%which, in turn, affects their multifractality.} 
%In 
%particular, we  conjecture that the original multifractal fingerprint 
%will be observable whenever these structures are large enough to 
%accommodate it and for resonances which are not too short-lived. We support 
%our conjecture using extensive numerical simulations.  

Our work is organized as follows. In Sec.~\ref{meth} we outline the methods used to 
measure multifractality and to describe open quantum maps.
In Sec.~\ref{model} we present the model, and in Sec.~\ref{sectheo} we 
provide the semiclassical theory for open quantum maps, which we then use to state
our conjecture concerning multifractality of resonance eigenstates. To  
support it, we present extensive numerical results in Sec.~\ref{secnum}. 
Finally, we summarize and discuss our findings in Sec.~\ref{secconc}.

%************************************************************************************************************************
%************************************************************************************************************************
\section{Methods}
%************************************************************************************************************************
\label{meth}
%************************************************************************************************************************
%************************************************************************************************************************
\subsection{Moment scaling and multifractality}
%************************************************************************************************************************
\label{methodmultif}
Consider a quantum state $\ket{\psi}$ belonging to a Hilbert space of dimension $N$. Multifractality of $\ket{\psi}$ expanded over a certain basis as $\ket{\psi} = \sum^{N-1}_{j=0} \psi_j \ket{j}$ can be computed by considering the measure defined by the norm-squared components $\mu_j=|\psi_j|^2$ and its associated moments $P_q = \sum_j \mu_j^q$ (with $q\in \mathbb{R}$). The scaling of $P_q$ with $N$ yields the multifractal dimensions $D_q$. Alternatively, one may divide the system (considered one-dimensional and of length $N$) into boxes of length $n$, and define the coarse-grained measure \mbox{$\mu_i(n)=\sum_{j=ni}^{n(i+1)-1}\mu_j$} and the corresponding coarse-grained moments $P_q(n)= \sum_i \mu_i(n)^q$. The multifractal (box-counting) dimensions $D_q$ are then defined \cite{falconer} through the scaling
{
	\begin{equation}\label{dq}
	P_q(n) \sim \left(\frac{n}{N}\right)^{(q-1)D_q}\, , \quad n/N \rightarrow 0
	\end{equation}
}
keeping $N$ fixed and varying the box size $n$. 

The exponents $D_q$ allow us to assess the way in which a state is distributed over the representation at hand. For extended states, $D_q$ is a constant for all $q$ with its value coinciding with $D_0$ (which is always equal to the dimension of the support, here $D_0=1$), whereas $D_{q>0}=0$  for localized states. If $D_q $ is a non-trivial function of $q$, the state is multifractal.
%, however, the state presents fluctuations which are neither restricted to a single order nor to a single scale then $D_q$ will be a nontrivial function of $q$, and consequently the distribution of the state will have an infinite number of independent scaling exponents, that is, it will be multifractal.
	
In practice, since a given value of $D_q$ as defined above demands a stable enough behavior through many scales in order to be numerically well defined, it is convenient to introduce a local multifractal dimension 
{
	\begin{equation}\label{dqn}
	\widetilde{D}_q(n) = \frac{1}{q-1}\log_{2} \left[P_q(2 n)/P_q(n)\right],
	\end{equation}
}	
	%
	%ALTERNATE FORM
	%
	\begin{comment}
	\begin{equation}\label{Dqn}
	\tilde{D}_q(n) = \frac{\log P_q(n+1)- \log P_q(n)}{\log (n+1) - \log n}
	\end{equation}
	\end{comment}
so as to
have a refined notion of what is happening from one scale to the next, helping thus determine if such scale-invariance is present or not. This will be the quantity we use to quantify multifractality in our system. More precisely, we will be interested in ranges of $n$ over which $\widetilde{D}_q(n)$ does not vary considerably, as this will indicate the range of scales over which a given state has a self-similar structure.

%To this end, we introduce a local multifractal dimension which we define as
	%
	%

%************************************************************************************************************************
\subsection{Open maps}\label{sec:gen}
%************************************************************************************************************************

\subsubsection{Quantum maps}\label{qmaps}
We consider a quantum map $U$ that is the quantum version of some classical automorphism of the 2-torus $\mathbb{T}^2$, denoted $\textbf{M}$. 
%(this classical map may or may not result from the discretization of a Hamiltonian flow). 
The corresponding phase space is usually represented as a square with periodic boundary conditions. Upon quantization, this periodicity yields a discrete Hilbert space of dimension $N$ with an associated effective Planck constant $h=1/N$ and a semiclassical limit \mbox{$N\to\infty$}. Position and momentum eigenbases can be denoted as $\{\ket{x_i}\}_{i=0}^{N-1}$,  $\{\ket{p_i}\}_{i=0}^{N-1}$ with $x_i,p_i\in \{0,1/N,\ldots,(N-1)/N\}$. A quantum map on the torus can then be expressed in either of these bases as an $N\times N$ unitary matrix $U$.
%
%\blue{In this section we expose some general properties of open maps...}
%

%**********************************************************************************
\subsubsection{Quantum opening}\label{qop}
%**********************************************************************************
Quantum mechanically, the opening in phase space can be achieved by means of a projection operator $\Pi$ onto the opening  \cite{novaes}. The projection operator may act before or after the quantum map operator, or both (as  in \cite{schomerus}). Here we choose to define the quantum evolution operator for the open system as 
\begin{equation}\label{oqmap}
\widetilde{U} = U (1-\Pi) \, .
\end{equation}
The projection operator can be represented in a specific basis $\{\ket{\xi_i}\}_{i=0}^{N-1}$ as
\begin{equation}\label{proj}
\Pi \equiv\sum_{i=0}^{\lfloor N\ell\rfloor-1} \ket{\xi_i}\bra{\xi_i}\, ,
\end{equation}
where $\ell\in (0,1)$ is a fixed number and $\lfloor N\ell\rfloor$ denotes the integer part of $N\ell$. The operator $(1-\Pi)$ in \eqref{oqmap} acts on a state \mbox{$\ket{\Psi}=\sum_j \Psi_j \ket{\xi_j}$} by setting its components \mbox{$\Psi_i, 0\leq i\leq \lfloor N\ell\rfloor-1$}, to zero. 

Since $\tU$ is not normal ($\tU\tU^\dagger \ne \tU^\dagger\tU$), it is necessary to distinguish between left $\ket{\Psi_j^-}$ and right $\ket{\Psi_j^+}$ eigenstates of the map, defined by $\widetilde{U} \ket{\Psi_j^+}={\lambda_j} \ket{\Psi_j^+}$ and $\bra{\Psi_j^-} \widetilde{U} = \bra{\Psi_j^-} {\lambda_j}$. As for the eigenvalues, they have the form $\lambda_j = e^{i E_j}e^{-\Gamma_j/2}$ (with $E_j,\Gamma_j$ real). Thus $\Gamma_j$ determines the spectral norm $\exp(-\Gamma_j/2)$ and the decay rate of the corresponding eigenstate. States with $|\lambda_j| \simeq 0$ at large $N$ correspond to short-lived states, which decay instantly upon iteration of $\widetilde{U}$. States with non-vanishing eigenvalues at large $N$ correspond to stable states. Among these last, states with $|\lambda_j| \simeq 1$ correspond to so-called supersharp resonances \cite{novaes2012}, and they are suspected to be generically tied to non-escaping periodic orbits \cite{borgonovi1991,wisniacki2008,novaes2009, pedrosa2012}. We will see that, in our system, the presence of classical periodic orbits that never reach the hole will indeed lead to the existence of this type of state.

%************************************************************************************************************************
\subsubsection{Classical opening}\label{classopen}
%************************************************************************************************************************
Classically, a region $\Omega=\{(x,p): 0\leq \xi\leq \ell\in[0,1]\}$ (with $\xi$ either $x$ or $p$) of phase space is defined as the opening by ceasing to propagate all trajectories of the classical map $\textbf{M}$ that fall into it. Let $\Omega_{m}=\textbf{M}^m\Omega$ be the iterates of the opening $\Omega$ under the classical map. One can define the regions \cite{keating}
\begin{align}
\label{Rmp}
\rrp &= \{z\in \Omega_{-m}: z\notin \Omega_{-n} \text{ for } 0\leq n < m\} ,\quad R_+^0 = \Omega \,,\\
\rrm &= \{z\in \Omega_{+m}: z\notin \Omega_{+n}  \text{ for }  1\leq n < m\} ,\quad R_-^1 = \Omega_1 \,.
\label{Rmm}
\end{align} 
The set $\rrp$ is the set of points which escape after exactly $m$ iterations of $\textbf{M}$, while $\rrm$ is the set of points that escape after exactly $m$ iterations of $\textbf{M}^{-1}$. The slight asymmetry between definitions \eqref{Rmp} and \eqref{Rmm} reflects the asymmetry in the definition \eqref{oqmap} of $\widetilde{U}$ (the projector acts \emph{before} the map).

Under successive iterations of $\textbf{M}$, a given point either escapes after exactly $m$ iterations or never escapes. If we denote by $K_+$ the forward-trapped set, that is, the set of points that never escape in the future, then phase space can be partitioned by $K_+$ and the $\rrp$, according to the minimum time it takes for points to reach the opening under iterations of $\textbf{M}$. In a similar way one defines the backward-trapped set $K_-$ of points that never escape in the past (under iterations of $\textbf{M}^{-1}$), and one can partition phase space by $K_-$ and the $\rrm$. The classical trapped set $K_0=K_+\bigcap K_-$ is the set of points that never escape in the future or the past. The sets $K_+$ and $K_-$ are generally fractals whose dimensions serve to relate properties of the open system with those of the original closed system (e.g.~through the Kantz-Grassberger relation \cite{kantz}), as well as to provide links to  properties of the associated quantum map as in, for example, the fractal Weyl law (FWL) \cite{lu,schomerus}.

%************************************************************************************************************************
%************************************************************************************************************************
\section{{The model: intermediate map}}
%************************************************************************************************************************
%************************************************************************************************************************
\label{model}
%************************************************************************************************************************
\subsection{Definition}
%************************************************************************************************************************
The model we study is obtained by quantization of a classical kicked system with discrete time dynamics on the 2-torus generated by
	\begin{equation}\label{cham}
	H(x,p,t) = p^2 -\gamma \{x\}\sum_n \delta(t-nT)\, ,
	\end{equation}
where $(x,p)$ is the phase space coordinate, $\{x\}$ denotes the fractional part of $x$ and $T$ is the time between successive kicks. The classical evolution, integrated over one period $T$, is then given by the map 
	\begin{equation}\label{cmap}
		\begin{array}{ccl}
		p_{n+1} &=& p_n + \gamma \\
		x_{n+1} &=& x_n + 2p_{n+1} 
		\end{array}
	\quad(\text{mod } 1)\, ,
	\end{equation}
which can be seen to be a combination of a kick in $p$ followed by free motion in $x$. 
The quantization of \eqref{cmap} over one period leads to a unitary evolution operator $U$ acting on a Hilbert space of dimension $N=1/{2\pi \hbar}$ \cite{giraud2004} which is given in momentum representation by
	\begin{equation}\label{qmap}
	U_{pp'} = \frac{e^{i \phi_p}}{N}\frac{1-e^{2\pi i N \gamma}}{1-e^{2\pi i(p'-p+N\gamma)/N}} \, ,
	\end{equation}
with $\phi_p=-2\pi p^2 /N$. Although this choice of $\phi_p$ is the proper one for the quantized map, one can instead take these phases as being randomly distributed in $[0,2\pi]$ \cite{bogomolny}, allowing for the possibility of constructing a random ensemble out of \eqref{qmap}. 

%************************************************************************************************************************
\subsection{Properties}
%************************************************************************************************************************
The properties of the classical map are tuned through the parameter $\gamma$ which can render the system either pseudointegrable or ergodic. Pseudointegrable systems are systems in which motion is restricted to $N$-dimensional surfaces  for \mbox{$N$-degree}-of-freedom systems, but the surfaces are more complicated (of higher genus) than for integrable systems \cite{richens1981}.
More precisely, for rational $\gamma=a/b$ the map \eqref{cmap} corresponds to an interval-exchange transformation in which motion is restricted to a union of $b$ 1-tori $\{(x,p): p=p_0+k\gamma\}_{0\leq k \leq b-1}$. On the other hand, for irrational $\gamma$ the map becomes ergodic, although still not mixing \cite{giraud2004}. These properties are in turn reflected on the properties of the quantum system. For $\phi_p$ uniformly distributed in $[0,2\pi]$ and $\gamma$ irrational, the statistics of the eigenphases of $U$ follow random matrix theory (RMT) for either the circular orthogonal ensemble or the circular unitary ensemble (depending, respectively, on whether the symmetry constraint $\phi_p = \phi_{N-p}$ is imposed or not), while the eigenstates of $U$ are extended in momentum space. On the other hand, for rational $\gamma=a/b$ the spectral statistics is intermediate between Poisson and RMT \cite{bogomolny,Bogomolny:2009gp}, and eigenstates are multifractal in momentum representation. The multifractality of its eigenfunctions and evolved wavepackets have been studied in \cite{MGG,GarciaMata:2009fk,MGGG,MGGG12}. In particular, it has been found that the strength of this multifractality decreases for larger $b$ (for instance, the information dimension $D_1$ behaves as $1-1/b$).

%************************************************************************************************************************
\subsection{Dynamical vs. random system}
%************************************************************************************************************************
\label{dynvsrand}
 There is an important difference between the dynamical map \eqref{qmap} with phases $\phi_p=-2\pi p^2/N$ and its random version where the $\phi_p$ are taken as random variables  uniformly distributed in $[0,2\pi]$. For the dynamical system, the phases $\phi_p$ are directly related to the kinetic energy operator, and the system possesses a well-defined classical limit. Semiclassical approximations will thus be in order. However, the main disadvantage is that a statistical treatment is then severely limited, as no ensemble averaging is possible.

The random-phase model \cite{bogomolny, MGG, MGGG} provides the possibility to have statistically significant results, and it retains some important features of the dynamical system. In particular, its multifractal properties are known to be very similar to the ones of the dynamical map for the closed system. Physically, it can mimic an average over quasimomenta that is present in experimental results for such maps. 
On the other hand, as in the random case there is no connection between the random phases $\phi_p$ from a disorder realization at a given $N$ to a realization at another $N$, its behavior might be expected to differ from the dynamical case in the semiclassical limit.

Although our theoretical framework will in principle concern only the case in which the phases are the dynamical ones, we will exploit the random-phase alternative as well, not only to have a setting in which we can perform significant statistics but also to probe a regime where there is no quantum-to-classical crossover of the escape dynamics \cite{schomerus, schomerus2005}, i.e. the open system is always in the quantum regime (see Sect. \ref{sec:xop_random}). As we shall see, this allows us to factor out some of the classical structures that would otherwise overlay the quantum multifractal behavior in which we are interested. {In particular, we will see that the multifractal properties of the random open map become increasingly different from those of the dynamical open map as the latter tends to the semiclassical limit}.
%************************************************************************************************************************
%************************************************************************************************************************
\section{Semiclassical structure and multifractality of open quantum systems: theory}
%************************************************************************************************************************
%************************************************************************************************************************
\label{sectheo}
%***********************************************************************************************************************
In this section, we first set out to describe the existing framework that accounts for the semiclassical structure of resonance eigenstates. We then formulate a conjecture which allows us to predict at which scales we may expect multifractality to manifest itself in the open system.
%
%%%
\subsection{Semiclassical structure}
\label{keatingconj}
%************************************************************************************************************************
We first consider some general properties concerning the semiclassical structure of the eigenstates of open maps such as $\widetilde{U}$.
We restrict ourselves to left eigenstates, the discussion for the right eigenstates being largely similar, with only  minor differences (at least for our ends) stemming from the fact that $\widetilde{U}$ has been defined asymmetrically, that is, with the complementary projector $(1-\Pi)$ acting only on the right. Our discussion follows mainly reference \cite{keating}.

Let $\ket{\Psi_j^-}$ be a left eigenstate of $\widetilde{U}$ with eigenvalue $\lambda_j$. The Husimi function of $\ket{\Psi_j^-}$ is defined as the normalized Weyl symbol of the density matrix $\rho\equiv\ket{\Psi_j^-}\bra{\Psi_j^-}$,
\begin{equation}
\label{husimifct}
\calH(z)=\frac{1}{\langle z|z\rangle}{\rm Tr}\left({\ket{z}\bra{z}\rho}\right).
\end{equation}
Here the $\ket{z}\equiv\ket{x,p}$ are coherent states of the harmonic oscillator, periodized on the 2-torus (see Appendix \ref{appcohstates} for a precise definition). The Husimi function $\calH(z)$ provides a convenient phase space representation of $\ket{\Psi_j^-}$ at scales above $\sim\sqrt{\hbar}$ (see Eqs.~\eqref{cohstates1}--\eqref{cohstates2}). In other words, 
%as coherent states of the harmonic oscillator are Gaussians of width $\sim\sqrt{\hbar}$ (see Eqs.~\eqref{cohstates1}--\eqref{cohstates2}), so that 
the Husimi function \eqref{husimifct} 
of the eigenvector %of length $N$ 
does not resolve structures at scales smaller than $\sim\sqrt{\hbar}=1/\sqrt{2 \pi N}$.

Our starting point is the assumption that the iterated action of the quantum propagator on a coherent state parallels the classical evolution so long as the number $m$ of iterates is smaller than some characteristic time $t_E$. The time $t_E$ is called the Ehrenfest time and is defined in the context of open quantum maps as the time it takes for a coherent state to stretch to the size of the opening. In the semiclassical limit we have $t_E\rightarrow \infty$, and the behavior of the open quantum map can be retraced to that of the classical system. The opposite limiting case $t_E\rightarrow 1$ marks a regime where the classical evolution becomes irrelevant and the escape dynamics is purely quantum. 

The \emph{quantum-to-classical crossover} regime corresponds to a finite number of iterates $m<t_E$ \cite{schomerus, schomerus2005}. In this regime, one can estimate how the Husimi function is distributed in phase space. Following the ideas of \cite{keating}, we now state two properties describing the support of the Husimi function and its relative weight on the classical structures described in the previous section. We relegate their detailed derivation to Appendix \ref{relativew}. 

On the one hand, one can show that whenever $\lambda_j\neq 0$ the Husimi function $\calH(z)$ is such that
\begin{equation}\label{keating_1}
\calH(z)\simeq 0 \quad\textrm{whenever}\quad z\notin K^{t_E}_{+}=\left(\bigcup_{0\leq m < t_E} R_+^m\right)^c \, ,
\end{equation}
where $^c$ denotes the complement with respect to the torus $\mathbb{T}^2$. In other words, this means that $\calH(z)$ is significant only for \mbox{$z\in K^{t_E}_{+}$}. This result holds only in the case in which $\lambda_j\neq 0$: for $|\lambda_j|$ close to zero, strong leakage outside of $K_+^{t_E}$ can occur. The set $K_+^{t_E}$ is the set of points that escape in more than $t_E$ iterations of $\boldmath{M}$, and can therefore be thought of as a finite-time approximation to the forward-trapped set $K_+=K_+|_{t_E\rightarrow \infty}$. 

On the other hand, one can consider the relative weight of the Husimi function of a given eigenstate on each of the $R_{-}^m$. It turns out that these weights can be related to the norm of the associated eigenvalue through 
\begin{equation}\label{keating_2}
\langle\Psi_j^- |\Pi_-^m| \Psi_j^- \rangle \approx
|\lambda_j|^{2(m-1)}(1-|\lambda_j|^2) \, ,
\end{equation}
for $1\leq m < t_E$, where $\Pi_{-}^m$ is the quantum projector onto $R_-^m$ (see \eqref{defproj} for its definition). The two results  \eqref{keating_1} and \eqref{keating_2} tell us that semiclassically the Husimi function concentrates on $K^{t_E}_{+}$, and that within this set its repartition over the classical regions $R_-^m$ is given by \eqref{keating_2}.

In essence, these results are statements about the Husimi function $\mathcal{H}(z)$ of a certain eigenstate on phase space cells $z$ of area $\hbar/2$ belonging to a certain region $R_-^m$. In this sense, it is important to note that they will be valid only for those $z$ that are sufficiently far from the borders of the region $R_-^m$ to which they belong. In other words, it must be kept in mind that \eqref{keating_1} and \eqref{keating_2} say little or nothing about points $z$ that are sufficiently close to the borders of a given region $R_-^m$ or, more precisely, those $z\in R_-^m$ for which $|\langle w|z\rangle|^2$ with $w\notin R_-^m$ is non-negligible. In the semiclassical limit, the size of the coherent states goes to zero and the contribution from the boundary becomes negligible. However, multifractal analysis for the open system requires that we fix the system size $N$ (and thus $\hbar$) and consider structures at various scales. Altogether, this means that whenever the relevant classical regions have characteristic sizes of the order of (the fixed) $\hbar$, we may have to consider possible ways of adapting the existing framework in order to contemplate finite-$\hbar$ effects and account for the eventual deviations from the theory outlined above. The relevance of these remarks will become evident when we come to the discussion of our numerical results.

%************************************************************************************************************************
\subsection{Multifractality of open quantum systems}\label{multifop}
%************************************************************************************************************************
The theory exposed above describes the left resonance eigenfunctions as having a semiclassical structure governed by the sets $R_-^{m}$ ($m<t_E$) and $K_+^{t_E}$. For open chaotic systems in the semiclassical limit, it is by now a fairly well established idea that resonance eigenstates $\ket{\Psi_j^{\pm}}$ should be \emph{on average}, and modulo quantum fluctuations, uniformly distributed over appropriate substructures partitioning $K_{\mp}$ \cite{casati,keating,ketzmerick}. It would be interesting to draw a parallel between the case of open chaotic systems and the case of open pseudointegrable systems, in order to use the semiclassical structure theory of the previous section to say something about the possibility of multifractality for resonance eigenstates. Before attempting to do so, however, there are some fundamental differences between the two settings that must be noted.

First, quantum fluctuations are important when discussing the multifractal structure of eigenstates. Thus, we have to be careful when we talk about \emph{average} properties. In our case, the wavefunction moments $P_q$ are calculated for \emph{individual} wavefunctions and then averaged over, as opposed to, for instance, averaging wavefunction components and then calculating the moments. This type of averaging would be akin to that employed in \cite{ketzmerick} when discussing the semiclassical structure of resonant eigenstates. There, the eigenstate Husimi functions are averaged over a small decay rate window.

Second, quantum multifractality is not a property of a \emph{phase space} distribution; rather, it concerns the distribution of wavefunction components in a particular basis. In situations in which the classical structures are sufficiently intricate in phase space, or rather badly oriented with respect to the basis where multifractality can be seen, it may be difficult to observe the multifractality of an eigenstate in such phase space regions. 

{Having commented on these differences, we conjecture that in the case of an open pseudointegrable system for which the eigenstates of the closed system are multifractal, resonance eigenstates should be multifractal on each of the substructures partitioning $K_{\mp}$, and that such multifractality will be observable only when the set $K_{\mp}$ and the sets $R_-^{m}$ allow for it.
The considerations of the previous section on the Husimi functions allow us to make a few general remarks concerning the substructures on which multifractality can be seen for the left eigenstates $\ket{\Psi_j^-}$.} Using \eqref{keating_2}, we distinguish between three regimes, depending on whether $\lam$ is close to 1, close to 0 (but different from it) or intermediate between the two. 

{\it In the case $\lam\simeq 1$}, apart from supersharp resonance states \cite{novaes2012} concentrated on non-escaping periodic orbits, Eq.~\eqref{keating_2} implies that states will have an almost vanishing weight on all regions $R_-^m$ with $m<t_E$ and will thus concentrate on the set $K_-^{t_E}$ defined as
\begin{equation}\label{eq:keating_3}
K^{t_E}_{-}=\left(\bigcup_{1\leq m < t_E} R_-^m\right)^c \,.
\end{equation}
Together with \eqref{keating_1} this implies that they will concentrate on $K_0^{t_E}\equiv K^{t_E}_{-} \bigcap K^{t_E}_{+}$, which at large $t_E$ goes to $K_0$. 

{\it In the case where $\lam$ is intermediate between 0 and 1}, states have their weight spread over many regions $\rrm$. Moreover, by virtue of \eqref{keating_1}, they will in fact be concentrated in the intersections $\rrm\bigcap K_+^{t_E}$. 

{\it Finally, when $\lam$ is slightly greater than 0}, states should be supported almost exclusively on $R_-^{1}\bigcap K_+^{t_E}$.

Multifractality in each of these cases will depend on the nature of the semiclassical structures (strongly dependent on $t_E$) and the scale at which one observes. In other words, we expect that the typical size of these structures will impose characteristic scales below which multifractality could in principle manifest.

\begin{figure} [!t]  %%% fig1
\includegraphics[scale=1]{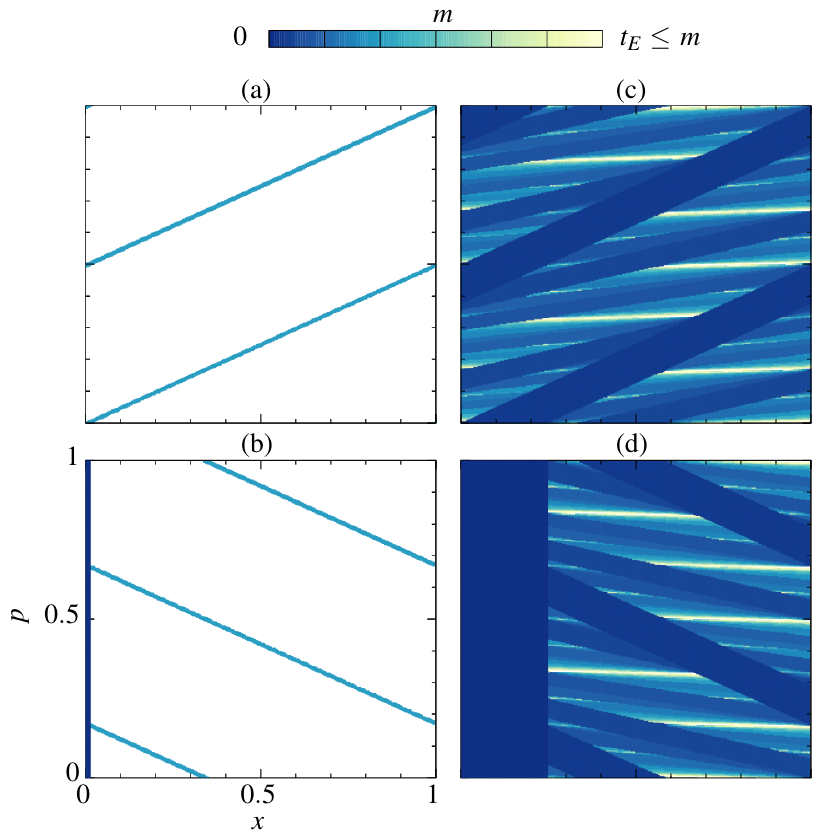}
\caption{(Color online) Regions $R_-^{m}$ (top row, $1\leq m< t_E$) and $R_+^{m}$ (bottom row, $0\leq m< t_E$) for the classical map opened in $x$ with $\ell=2^{-6}$ (left column), $2^{-2}$ (right column), and $\gamma=1/3$. The color scale indicates the number of iterates $m$. White regions include all those regions $R_{\pm}^{m}$ corresponding to $m\geq t_E$ ($t_E=2$ for $\ell=2^{-6}$ and $t_E=28$ for $\ell=2^{-2}$), i.e. the sets $K_-^{t_E}$ (top row) and $K_+^{t_E}$ (bottom row) whose intersection gives the trapped set $K_0^{t_E}$. As $t_E\rightarrow\infty$ the sets $K_\pm^{t_E}$ tend in fact to (the same) horizontal connected sets of periodic points which stay in the system forever, either under backward (top) or forward (bottom) evolution.}
\label{classical_xop}
\end{figure}

\subsection{Application to our model}

For a chaotic map with classical Lyapunov exponent $\nu$, a wavepacket of size $\sim\sqrt{\hbar}$ reaches the size $\ell$ of the hole in a time $t_E\sim \-\ln \hbar/\nu$ (or, equivalently, $t_E\sim\ln N$). In our system, the classical evolution of a point $(x,p)$ under \eqref{cmap} is given by
\begin{equation}\label{cevo}
(x,p) \mapsto_{t \text{ iterations}}(x',p')=(x+2p \,t+\gamma\,t(t+1),p+\gamma\,t)\, ,
\end{equation}
so that a point $(x+\delta x,p+\delta p)$ will be mapped to a point \mbox{$(x'+\delta x',p'+\delta p')$} with $\delta x'=\delta x+2t\delta p$ and $\delta p'=\delta p$. Therefore under time evolution an initial domain of linear size $\sqrt{\hbar/2}$ around $(x,p)$ does not stretch in the $p$ direction but stretches as $(2t+1)\sqrt{\hbar/2}$ along the $x$ direction. In particular, this means that opening the map in the $p$ direction gives an Ehrenfest time $t_E\rightarrow\infty$, whereas for an opening in the $x$ direction an initial coherent state will stretch to the size $\ell$ of the opening in a finite number of iterations
\begin{equation}
\label{tehr}
t_E=\lceil \ell\sqrt{\pi N}  - 1/2\rceil \, ,
\end{equation}
where we have used $\hbar=1/(2\pi N)$. In short, this means that opening the dynamical quantum map in $x$ or $p$ yields the finite $t_E> 1$ and infinite $t_E$ regimes, respectively. As for the special case $t_E=1$, it can be probed by using the random-phase map opened in $x$ (see Section \ref{sec:xop_random} below).

%************************************************************************************************************************
%************************************************************************************************************************
\section{Numerical results}
%************************************************************************************************************************
%************************************************************************************************************************

%%%%%%%%%%%%%%%%%%
\begin{figure} [!t] %% fig2
\includegraphics[scale=1]{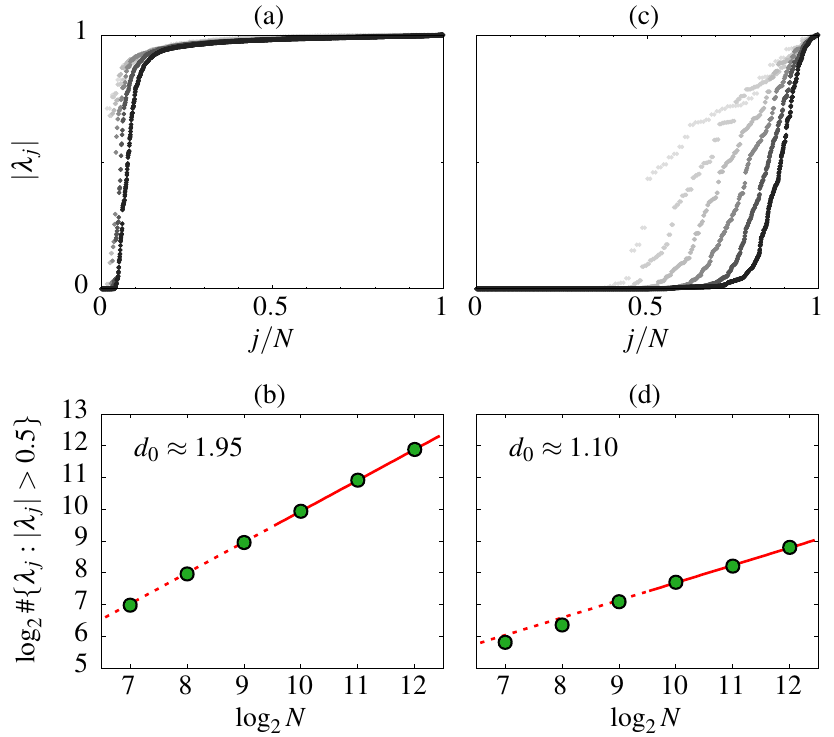}
\caption{(Color online) Spectral norms $|\lambda_j|$ (top row) of the dynamical map with $\gamma=1/3$, opened in ${x}$ with $\ell=2^{-6}$ (a) and $2^{-2}$ (c), for various system sizes $N=2^{7},...,2^{12}$ (light to dark circles). The scaling of the number of resonances with norm above 0.5 is shown in the bottom row for each case $\ell=2^{-6}$ (b) and $2^{-2}$ (d). Circles correspond to the data, while the line is the linear fit (with the solid segment indicating the fitting range), whose slope corresponds to $d_0/2$.}
\label{spectral_norms_x_dynamical}
\end{figure}
%%%%%%%%%%%%%%%%%%%%%%%%%%%%%% 

%
\begin{figure} [!h]  %% fig3
\includegraphics[scale=1]{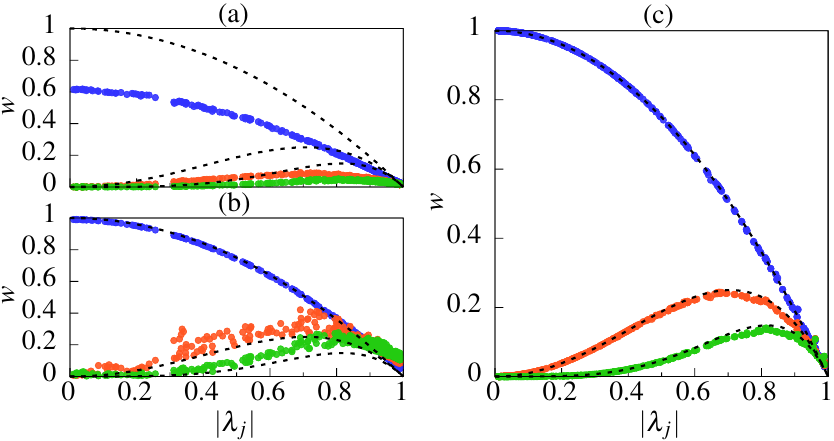}
\caption{(Color online) Numerically calculated weights $\langle\Psi_j^- |\Pi_-^m| \Psi_j^- \rangle $ of left eigenstates of the open dynamical map on regions $R_{-}^m$ for $m=1, 2$ and $3$ (blue, red, green; dark to light, resp.). Dashed black lines correspond to the theoretical prediction \eqref{keating_2} for each of these corresponding regions (from top to bottom in each panel). The map (with $N=2^{12}$ and $\gamma=1/3$) is opened in $x$ with $\ell=2^{-6}$ ((a) and (b)) and $\ell=2^{-2}$ (c). For $\ell=2^{-6}$, panel (a) corresponds to the weights calculated on the original regions (i.e. in the same way as for the large opening), while panel (b) corresponds to the weights on the broadened regions.}
\label{weightsRm_xop}
\end{figure}

\label{secnum}

%\subsubsection{Quantum states: semiclassical picture}\label{semiclassical}
%%%
%
%%

We will now present the results of our numerical investigations for the map Eq.~\eqref{qmap}. We will consider two 
types of openings defined by the choice of basis in the projector Eq.~\eqref{proj}: one in  position (Sec. \ref{sec:xop} and \ref{sec:xop_random}) and  the other in momentum (Sec. \ref{sec:pop}).
%open either in position or in momentum. 
We will illustrate our results for $N=2^{12}$ and two different opening sizes: a small opening $\ell=2^{-6}$,
for which Eq.~\eqref{tehr} yields an Ehrenfest time $t_E=2$, and  a large opening $\ell=2^{-2}$ with $t_E=28$.
When opening in position, we need to further make the distinction between the dynamical and random versions of the map, as discussed in \ref{dynvsrand}.

%************************************************************************************************************************
%************************************************************************************************************************
\subsection{Opening in position for the dynamical map}\label{sec:xop}
%************************************************************************************************************************
%************************************************************************************************************************
%%%%%%%%%%%%%%%%
Here we consider the case in which the opening is in $x$ and the quantum map is \eqref{qmap} with dynamical phases $\phi_p=-2\pi p^2 /N$. 
According to Eqs.~\eqref{oqmap}--\eqref{proj}, the open quantum map is defined as
\begin{equation}
\widetilde{U} = U\sum_{i=\lfloor N\ell\rfloor}^{N-1} \ket{x_i}\bra{x_i}\,.
\end{equation}
It is obtained in $x$ representation by taking the Fourier transform of \eqref{qmap} and setting the first $\lfloor N\ell\rfloor$ columns to 0. 
We will be interested in the properties of its eigenvalues $\lambda_j$ and its left eigenstates $\ket{\Psi_j^-}$. 
%************************************************************************************************************************

%%%%%%%%%%%%%%%%
%%%%%%%%%%%%%%%%
\begin{figure*} [!t] %fig4
\includegraphics[scale=1]{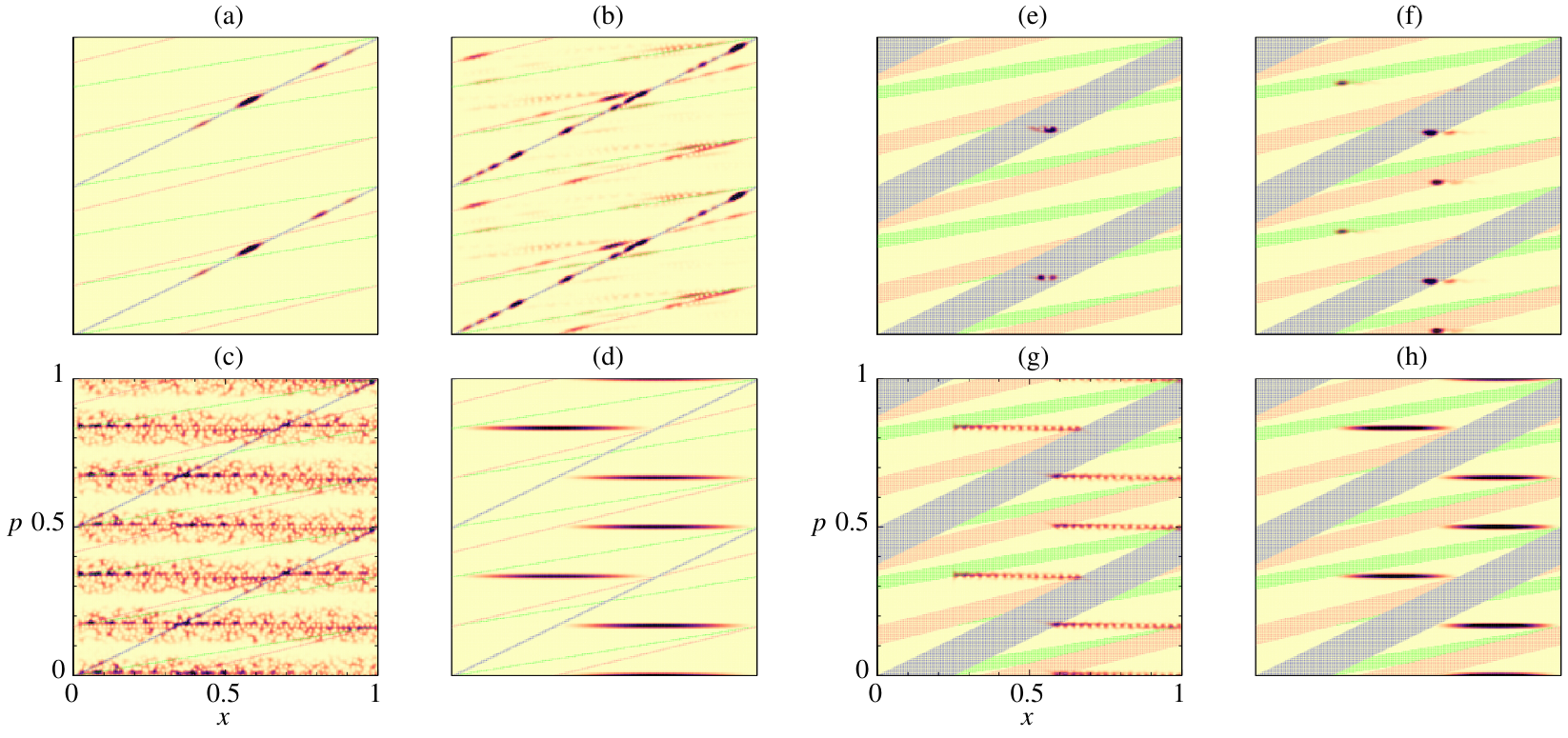}
\caption{(Color online) Husimi representation of four eigenstates of the dynamical map opened in $x$ for $\ell=2^{-6}$ (a-d) and $\ell=2^{-2}$ (e-h). Superimposed on each panel are the classical regions $R_-^1$, $R_-^2$ and $R_-^3$ (blue, red, green; dark to light, resp.). The eigenvalue norm $|\lambda|$ associated to each eigenstate is (a) 0.02223, (b) 0.71880, (c) 0.97335, (d) 0.99998, (e) 0.02007, (f) 0.72074, (g) 0.097331, (h) 0.99989.
}
\label{husimi_xop_dynamical}
\end{figure*}
%%%%%%%%%%%%%%%%
%%%%%%%%%%%%%%%%

\subsubsection{Classical properties}
\label{classpropdyn}
The classical structures described in the previous section find clear illustration in the present setting. 
They are displayed in \fig{classical_xop}. 
The shaded areas correspond to the regions $R_\pm^{m}$. In particular, in {\fig{classical_xop}(a)} the two stripes correspond to the set $R_-^1$ (which is the only  one displayed since $t_E=2$ for $\ell=2^{-6}$ and $N=2^{12}$), while the set $K^{t_E}_{-}$ corresponds to the white region and covers almost all of phase space. By contrast, in {\fig{classical_xop}(c) corresponding to $\ell=2^{-2}$ (where $t_E=28$)} the set $R_-^1$ corresponds to the two darkest stripes, while $K^{t_E}_{-}$ is reduced to six thin and almost one-dimensional white regions. Similarly, the set $K^{t_E}_{+}$ corresponds to the white regions in the bottom row {(Figs.~\ref{classical_xop}(b) and \ref{classical_xop}(d))}.

In the semiclassical limit $t_E\to\infty$, 
%both sets tend to the same horizontal stripes. 
the sets $K^{t_E}_{-}$ and $K^{t_E}_{+}$ become equal, and both tend to the trapped set $K_0$. For any $\ell$, this set of trapped points lies within $2b$ horizontal segments, as can be deduced from
the analysis of the classical map \eqref{cmap}. Indeed, as mentioned above, after $t$ iterations a point $(x,p)$ is mapped to $(x+2tp+\gamma\, t(t+1),p+t\gamma)$. It is then easy to see that in the closed map any point with a momentum of the form $p=k/(2b)$ with $0\leq k\leq 2b-1$ corresponds to a periodic point, with a period determined by its position $x$. In the open system, the trapped set $K_0$ thus corresponds to periodic points of the closed map whose trajectory never goes through the opening.

%

%************************************************************************************************************************
\subsubsection{Spectral properties}

As discussed in section \ref{sec:gen}, we expect that multifractality of eigenvectors of $\widetilde{U}$ will be governed by the classical structures and by its eigenvalues $\lambda_j$. In {Figs.~\ref{spectral_norms_x_dynamical}(a) and \ref{spectral_norms_x_dynamical}(c)} we display the spectral norms $|\lambda_j|$ ordered by increasing value of their modulus ($|\lambda_j| \leq |\lambda_{j+1}|)$ as a function of  the rescaled index $j/N$.

Some predictions can be made about the number of short-lived or long-lived states. On the one hand, the number of short-lived states (with eigenvalues $\simeq 0$) can be estimated by counting the number of independent minimal wavepackets whose escape time is smaller than the Ehrenfest time $t_E$. At the same time, the number of long-lived states can be estimated by the fractal Weyl law (FWL) \cite{lu}, a conjecture relating the asymptotics of the resonance distribution with the fractal dimension $d_0$ of $K_0$. More specifically, the FWL states that for an $N$-dimensional Hilbert space the number of resonances ${\lambda}$ such that $|\lambda|>r$ grows as $A(r)N^{d_0/2}$ where $A(r)$ is some function of $r$. In the case in which the Ehrenfest time is finite, we may expect that the FWL is governed by properties of the set $K_0^{t_E}$. We check these predictions in {Figs.~\ref{spectral_norms_x_dynamical}(b) and \ref{spectral_norms_x_dynamical}(d)}. For $\ell=2^{-6}$  ($t_E=2$, {Fig.~\ref{spectral_norms_x_dynamical}(b)})  we see that $d_0\approx 1.95$ so that these states are located on an almost two-dimensional support, something that is consistent with {Figs.~\ref{classical_xop}(a) and \ref{classical_xop}(b)}, where $K_0^{t_E}$ covers almost all of phase space. As $\ell$ (and thus $t_E$) increases, $d_0$ gets closer to $1$ {(Fig.~\ref{spectral_norms_x_dynamical}(d))}, in accordance with the fact that in this limit $K_0^{t_E}$ approaches $K_0$ {(see Figs.~\ref{classical_xop}(c) and \ref{classical_xop}(d))}, becoming almost one-dimensional.

%************************************************************************************************************************

%Top three
\begin{figure*} [!t]  %% fig5
\includegraphics[scale=1]{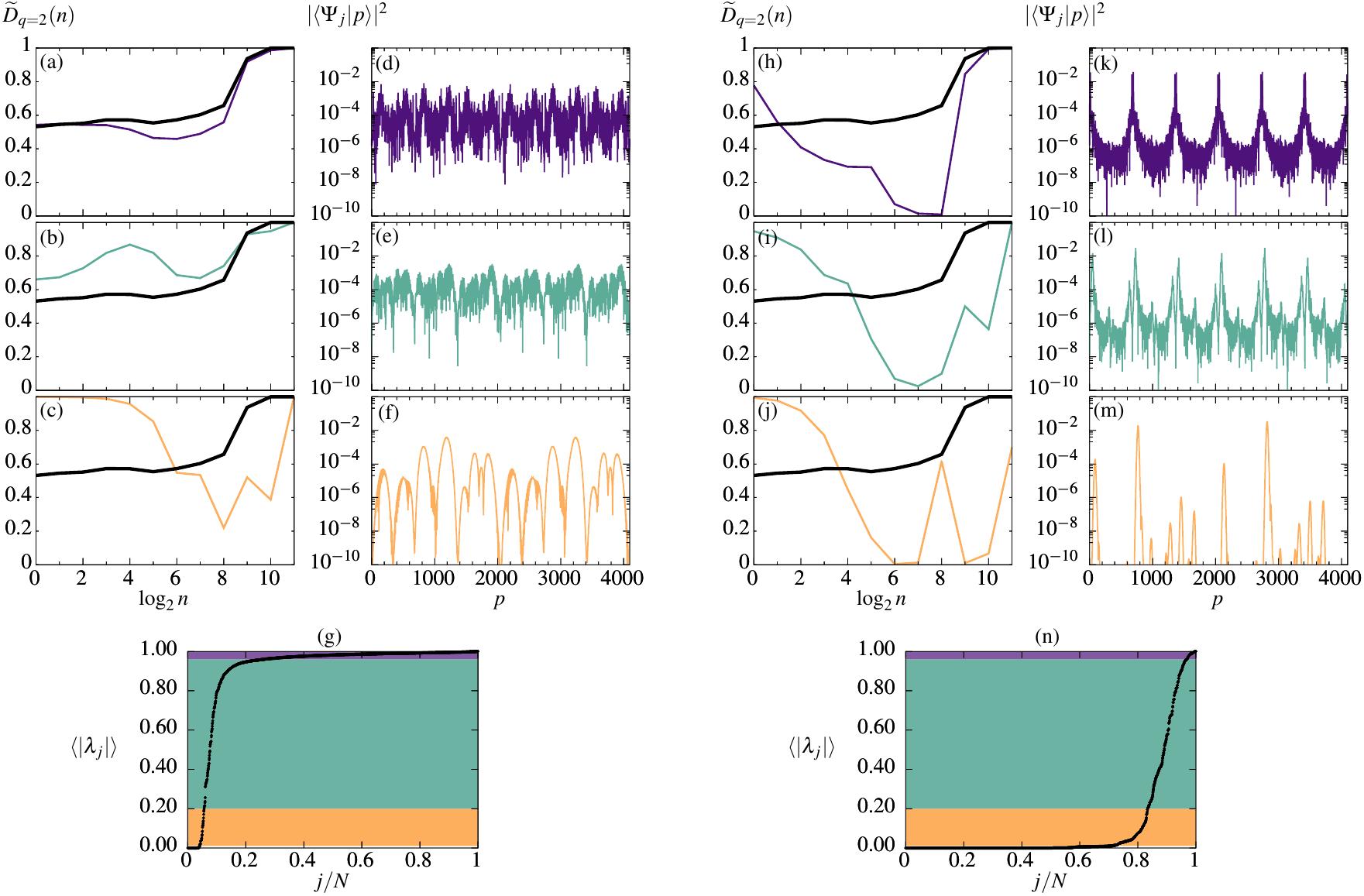}
\caption{(Color online) Multifractal analysis for the dynamical map opened in $x$ with $\ell=2^{-6}$ (a-g) and $\ell=2^{-2}$ (h-n). Top three rows in each case ($|\lambda_j|$ decreases from top to bottom): box-size $n$ behavior of $\widetilde{D}_2(n)$ for the left eigenstates of the open map $\ket{\Pl_j}$ (thin colored lines) and those of the closed map $\ket{\Pz_k}$ (thick black line); for $\ket{\Pz_k}$ the average is over the $N$ available states, whereas for $\ket{\Pl_j}$ we average over all those states (out of the $ N$) falling within a small norm range indicated by the color and shade of the curve in panels (g) and (n) displaying the spectral norms $|\lambda_j|$ for each case. To the right of each $\widetilde{D}_2(n)$ plot, we display $|\bra{\Pl_j}p \rangle|^2$ for an exemplifying state in the corresponding $|\lambda_j|$ window. Results correspond to system size $N=2^{12}$ and $\gamma=1/3$.}
\label{mf_xop_dynamical}
\end{figure*}
As explained in Section \ref{keatingconj}, eigenvectors of the dynamical map should be such that their Husimi function follows the general results \eqref{keating_1} and \eqref{keating_2}. According to \eqref{keating_1}, states should concentrate on the set $K^{t_E}_{+}$, and within this region the weight of the state on the sets  $R_{-}^m$ should be given by Eq.~\eqref{keating_2}. This prediction is checked in {Fig.~\ref{weightsRm_xop}} for the two openings $\ell=2^{-6}$ and $\ell=2^{-2}$. For the large opening ({Fig.~\ref{weightsRm_xop}(c)}), the relative weight in regions $R_{-}^m$ is quite accurately given by \eqref{keating_2} for all values of $\lam$ and  $m=1,2,3$. 
On the other hand, for the small opening ({Fig.~\ref{weightsRm_xop}(a)}) we see that the weights fall significantly below the theoretical prediction. To understand this behavior, we show
in {Figs.~\ref{husimi_xop_dynamical}(a-d)} the Husimi function of various eigenstates, and we observe that, at least for the shortest-lived states ({Figs.~\ref{husimi_xop_dynamical}(a) and \ref{husimi_xop_dynamical}(b)}), the Husimi function seems to be concentrated primarily on the $R_{-}^m$. At the same time, these regions consist of extremely thin stripes whose width turns out to be of order $\sim\sqrt{\hbar}$
\cite{footnote}.
%.
We thus expect that, even though  the Husimi functions are mainly on the $R_{-}^m$, they have non-negligible weight just outside the boundaries of these. In order to verify if this picture is accurate, we calculate the relative weight of the Husimi functions on the \emph{broadened} version $\widetilde{R}_-^{m}$ of the original regions, given by
\begin{equation}
\label{defproj}
\widetilde{R}_-^{m} = \{z\in \mathbb{T}^2: d(z,z')\leq 2\sqrt{\hbar/2} \text{ for all } z'\in R\},
\end{equation}
where $d$ is the Euclidean distance; in other words, this broadening amounts to a thickening of the original regions by two Planck cells. The results are shown in {Fig.~\ref{weightsRm_xop}(b)}. We see that the missing weight reappears and a good agreement with theory is attained, with appreciable deviations only for $m\geq t_E=2$.

These results show how one may take into account important finite-$\hbar$ effects when the possibility of going further into the semiclassical limit is not an option. Eventually, of course, if the relevant classical structures were to become much smaller than $\hbar$, the broadening procedure should not be expected to hold ground. 
\subsubsection{Quantum states: multifractality}
Multifractality of eigenstates of the open system can be analyzed in light of the description of the classical properties of the dynamical map given in Section \ref{classpropdyn}. 
%
%In \fig{husimi_xop_dynamical} we display the Husimi function of various eigenstates and 
In \fig{mf_xop_dynamical} we show the results of the multifractal analysis. The spectrum is divided into regions (given by the shaded areas in the spectrum displayed in {Figs.~\ref{mf_xop_dynamical}(g) and \ref{mf_xop_dynamical}(n)}), over which we average in order to determine the multifractal exponents associated with each region. We display results for $\widetilde{D}_{q}$ with $q=2$, for the open (thin colored curves) as well as for the closed map (thick black curve) in {Figs.~\ref{mf_xop_dynamical}(a-c) and \ref{mf_xop_dynamical}(h-j)}. Similar results were obtained for other values of $q$ (data not shown). We observe the following behavior depending on the value of $|\lambda_j|$.

For $\lam\simeq 1$, states should be concentrated on  $K_0^{t_E}$. For the small opening $\ell=2^{-6}$ (small $t_E$), this set is  a large part of phase space {(see Figs.~\ref{classical_xop}(a) and \ref{classical_xop}(b))}. 
As seen in {Fig.~\ref{husimi_xop_dynamical}(c)}, these states do not spread over the full available space. {Fig.~\ref{mf_xop_dynamical}(a)} shows that most of these
states are instead multifractal on this support. Nonetheless, we note that for $\lam$ extremely close to $1$, some states become localized on $K_0$ (a set consisting of continuous families of periodic points, contrary to the general chaotic case), and should remain so in the semiclassical limit. An example is shown in {Fig.~\ref{husimi_xop_dynamical}(d)}. Since the set $K_0$ is one-dimensional in $p$, {multifractality could only manifest itself at scales below $\sim\sqrt{\hbar}$}, but in fact these eigenstates are found to be localized on $2b=6$ basis states and display no multifractality. On the other hand, {Figs.~\ref{mf_xop_dynamical}(a) and ~\ref{mf_xop_dynamical}(d)} show that in the regime $\lam$ close to 1 most states are not of this type. 
For the large opening $\ell=2^{-2}$ (large $t_E$), $K_0^{t_E}$ is close to $K_0$, and hence all the long-lived states are of the extremely localized type ({see Figs.~\ref{husimi_xop_dynamical}(g) and \ref{husimi_xop_dynamical}(h), and Figs.~\ref{mf_xop_dynamical}(h) and \ref{mf_xop_dynamical}(k)}).

States with $\lam$ intermediate between 0 and 1 concentrate on $\rrm\bigcap K_+^{t_E}$. For $\ell=2^{-6}$, $K_+^{t_E}$ covers most of phase space, which means that states can, in principle, lie anywhere within $R_-^{m}$, mainly for small $m$. An example of such a state is shown in {Fig.~\ref{husimi_xop_dynamical}(b)}. The multifractal analysis of {Fig.~\ref{mf_xop_dynamical}(b)}, shows that these states present some multifractality at the smallest scales ($0\leq \log_2 n\leq 2$). However, the fact that these regions consist of oblique stripes whose projection in the $p$ direction is the entire interval $[0,1]$, makes it difficult to relate this multifractality to a given phase space structure. For $\ell=2^{-2}$, $K_+^{t_E}$ gets closer to $K_+$ and despite the fact that the $R_-^{m}$ are much larger than in the case $\ell=2^{-6}$, the intersections $\rrm\bigcap K_+^{t_E}$ are very small, especially in the $p$ direction (where multifractality is concerned). These features appear clearly in {Fig.~\ref{husimi_xop_dynamical}(f)}. The corresponding multifractal analysis of {Fig.~\ref{mf_xop_dynamical}(i)} shows that, on average, states in this regime do not present any evident multifractal structure.

States with $\lam\simeq 0$ concentrate on $R_-^1\bigcap K_+^{t_E}$. For the small opening (since $K_+^{t_E}$ is almost all of phase space) this means essentially all of $R_-^1$. {Fig.~\ref{husimi_xop_dynamical}(a)} shows that this type of state, although confined to $R_-^1$, does not spread over the whole available space. For the large opening, the intersection $R_-^1\bigcap K_+^{t_E}$ amounts to an almost zero-dimensional region [compare the white regions with the darkest regions in {Figs.~\ref{classical_xop}(c) and \ref{classical_xop}(d)}], something that is reflected in {Fig.~\ref{husimi_xop_dynamical}(e)}. Results of the multifractal analysis are displayed in {Figs.~\ref{mf_xop_dynamical}(c) and \ref{mf_xop_dynamical}(j)}. The behavior of $\widetilde{D}_q$ for both opening lengths with $\lam\simeq 0$ shows that there is no multifractality for either of these cases, not even at the smallest scales corresponding to regions \emph{within} $R_-^1\bigcap K_+^{t_E}$. In fact, the multifractal analysis shows that they manifest instead a small-scale ergodicity there. We will encounter a similar type of behavior as well when considering the case $\lam\approx0$ for the map opened in $p$.

It is interesting to note that multifractality in our system clearly depends on $t_E$, and thus on $\hbar$ (it is more visible for small $\hbar$), even though multifractality in the closed system is manifested independently of $\hbar$.

%************************************************************************************************************************
\subsection{Opening in position for the random map}\label{sec:xop_random}
%************************************************************************************************************************

\begin{figure}[!t]  %%fig6
\includegraphics[width=.99\linewidth]{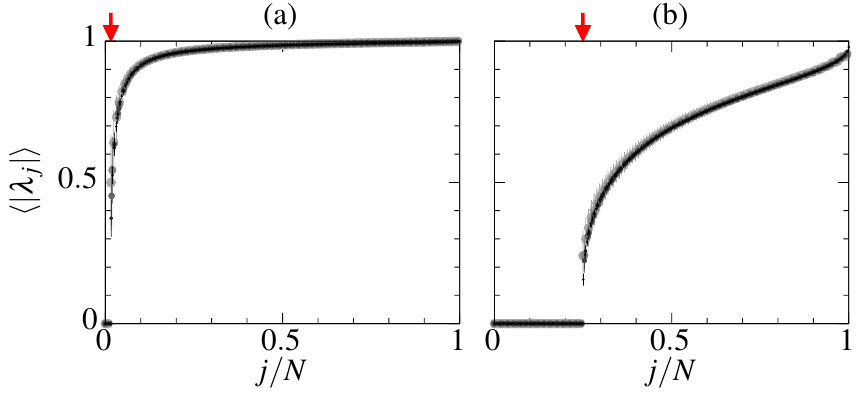}
\caption{(Color online) Average spectral norms $\langle |\lambda_j| \rangle$ of the random map for $\gamma=1/3$, opened in ${x}$ with $\ell=2^{-6}$ (a), $2^{-2}$ (b) and for various system sizes $N=2^{7}$ (lightest gray)$,...,2^{12}$ (darkest gray). For each $N$ we diagonalize $M=2^{17}/N$ random-phase realizations. The index \mbox{$j\in\{1,...,N\}$} is defined by the ordering $|\lambda_j| \leq |\lambda_{j+1}|$, and the average $\langle |\lambda_j| \rangle$ is taken over the $M$ realizations at fixed $j$. The arrow at the top of each plot marks the (rescaled) theoretical prediction $\ell$ for the size of the short-lived sector.}
\label{spectral_norms_xop_random}
\end{figure}

\begin{figure}[!t] %%fig7
\includegraphics[scale=1]{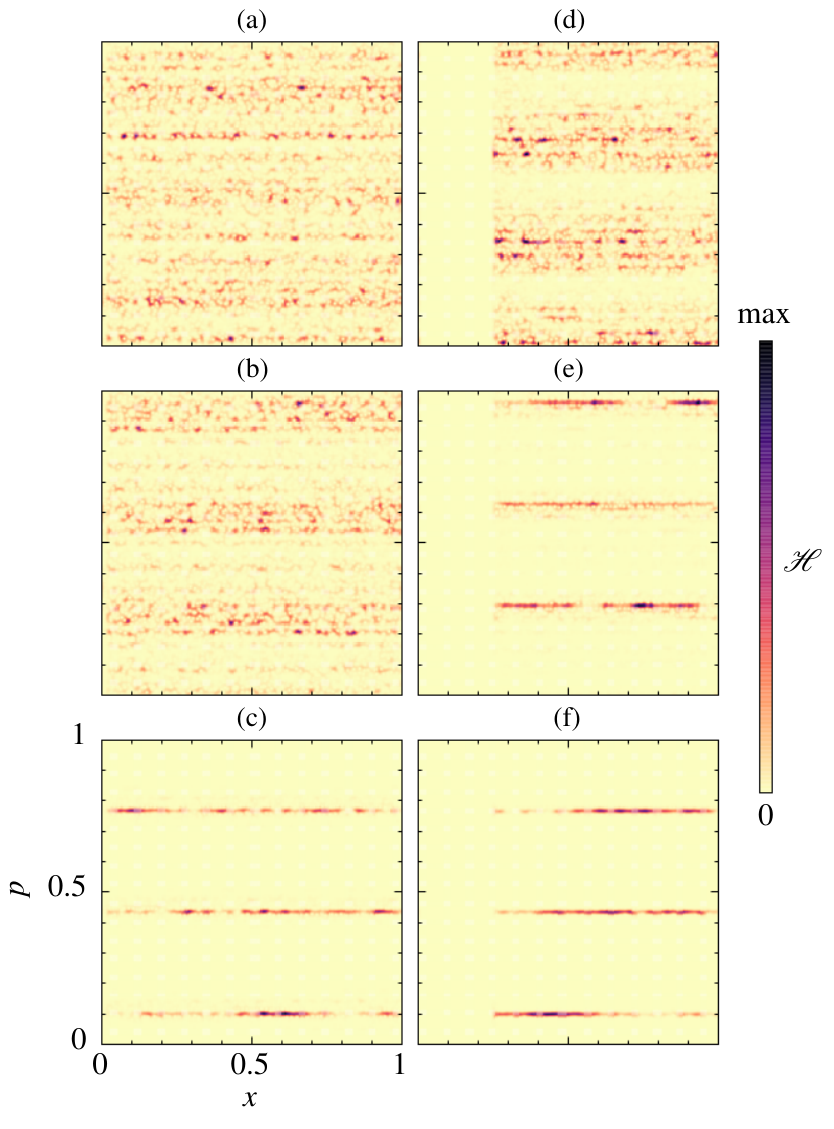}
\caption{(Color online) Husimi representation of three left eigenstates of a particular realization of the open random map with $\gamma=1/3$, \mbox{$N=2^{12}$} and opened in $x$ with $\ell=2^{-6}$ (a-c), $2^{-2}$ (d-f). The eigenvalue norm $|\lambda|$ associated to each eigenstate is (a) 0.740, (b) 0.900, (c) 0.995, (d) 0.661, (e) 0.750, (f) 0.949.}
\label{husimi_xop_random}
\end{figure}
%************************************************************************************************************************
\begin{figure*}[!ht] %%fig8
\includegraphics[scale=1]{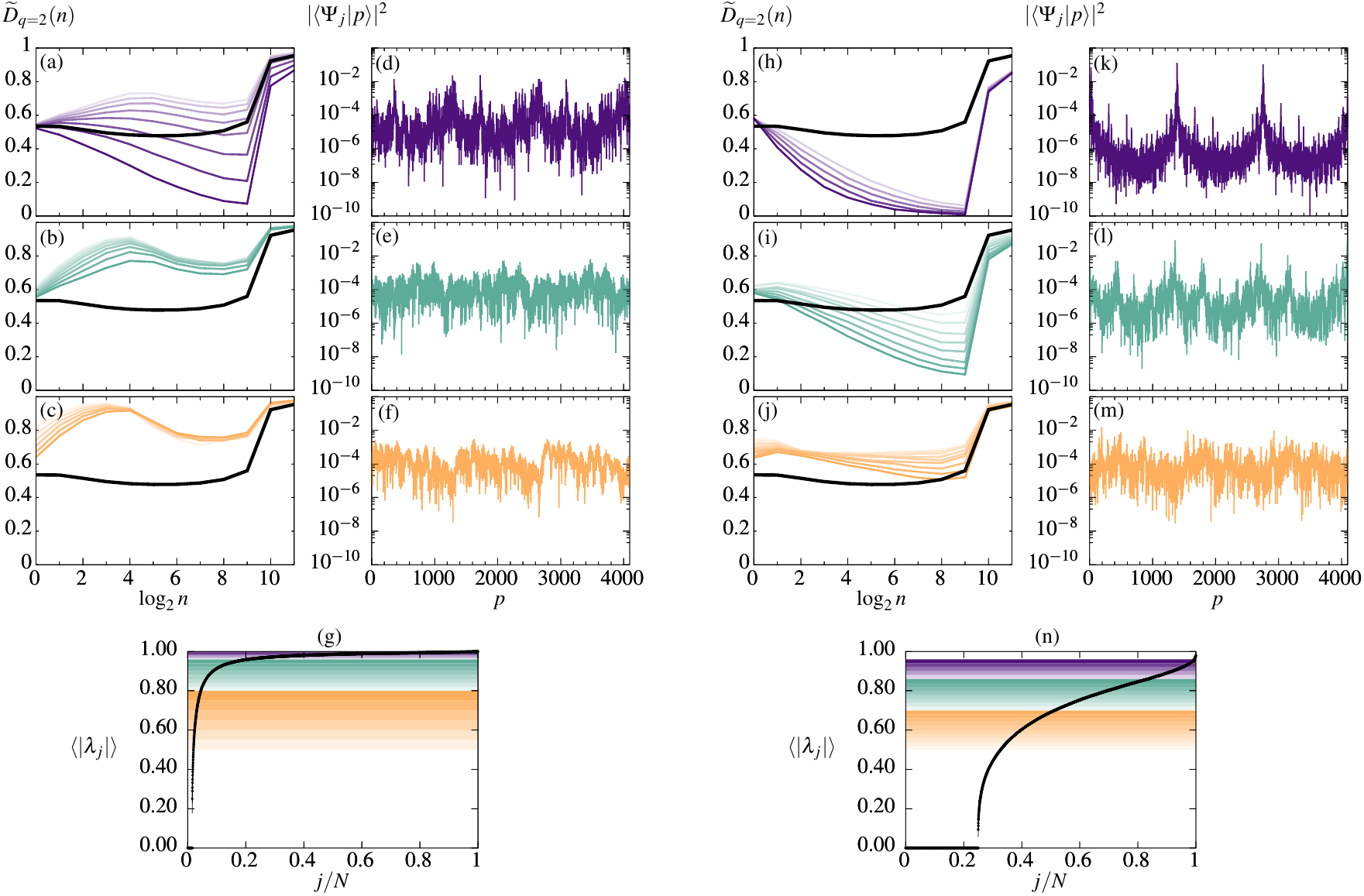}
\caption{(Color online) Multifractal analysis for the random map opened in $x$ with $\ell=2^{-6}$ (a-g) and $\ell=2^{-2}$ (h-n). Top three rows in each case ($\langle|\lambda_j|\rangle$ decreases from top to bottom): box-size $n$ behavior of $\widetilde{D}_2(n)$ for the left eigenstates of the open map $\ket{\Pl_j}$ (thin colored lines) and those of the closed map $\ket{\Pz_k}$ (thick black line); for $\ket{\Pz_k}$ the average is over the $M\times N$ available states, whereas for $\ket{\Pl_j}$ we average over all those states (out of the $M\times N$) falling within a small norm range indicated by the color and shade of the curve in panels (g) and (n) displaying the spectral norms $\langle|\lambda_j|\rangle$ for each case. To the right of each $\widetilde{D}_2(n)$ plot, we display $|\bra{\Pl_j}p \rangle|^2$ for an exemplifying state in the corresponding $\langle|\lambda_j|\rangle$ window. Results correspond to $M=32$ random realizations with $N=2^{12}$ and $\gamma=1/3$.}
\label{mf_xop_random}
\end{figure*} 

\subsubsection{Spectral properties}
\label{specrandom}
We now turn to the case in which the phases $\phi_p$ are taken as random variables. In this case there is no longer an underlying classical dynamics, although the system still retains some features of the dynamical map. In \eqref{qmap}, the dynamical phase $\phi_p$ corresponds to the free evolution operator $\exp(-i\hat{p}^2/\hbar)$, which in the $p$ representation is diagonal with entries $\exp(-2i\pi p^2/N)$. This free evolution term corresponds to the shift $2p$ along $x$ in the classical map \eqref{cmap}, which can be obtained from integration of the Hamilton equation. One can rewrite this equation as $\dot{x}=\partial(-\hbar \phi_p)/\partial p$ (where $\phi_p$ depends on $\hbar$, so that the whole equation is classical). In the random case, one may argue that this equation yields a classical evolution where $x$ is randomly kicked in $[0,1]$, while the evolution in $p$ remains as in the dynamical case. One way to check the validity of this picture is to  trace the short-time dynamics of cells in a phase space coarse-grained by coherent states. Computing the first iterates of a gaussian wavepacket under an instance of the random map, we found (data not shown) that it expands along $x$ to the size of the system in only one iteration, although an average over many first iterates (corresponding to different random instances of the map) was necessary in order to observe a completely uniform spread. The fact that an initial wave packet spreads to the size of the system in only one iteration implies that we are always in the quantum regime, i.e. $t_E=1$. Furthermore, this property is independent of $N$.

In \fig{spectral_norms_xop_random} we display the average spectral norms $\langle |\lambda_j| \rangle$. Since  $t_E=1$, the number of short-lived states $\langle |\lambda_j| \rangle\simeq 0$ corresponds to the number of wavepackets that escape immediately, that is, wavepackets supported on the opening. {Therefore the size of the short-lived sector should simply be $\lfloor N\ell\rfloor$ (indicated by an arrow at the top of \fig{spectral_norms_xop_random}; note the rescaling of the horizontal axis), which is in agreement with the form of the spectrum. As regards the long-lived sector, we first note that the set $K^{t_E}_{0}|_{t_E=1}$ is the entire phase space minus the opening, and hence a two-dimensional set.}
%
%Moreover, the set $K^{t_E}_{+}$ is the set of points which escape in more than $1$ iterations: this covers almost the whole phase space, so that its dimension is 2. Similarly the set $K^{t_E}_{-}$ of points which escape in more than $1$ iterations under the reverse map is two-dimensional. The intersection $K^{t_E}_{0}$ of the two sets, which is expected to govern the FWL, is therefore of dimension $d_0=2$.
%
The FWL then predicts that the number of long-lived resonances should grow as $\sim N$. This is precisely what is observed in \fig{spectral_norms_xop_random}, where points corresponding to different system sizes $N$ fall on the same curve after rescaling by $N$.

%, corresponding to the fact that there is no structure to be resolved in going to higher $N$.

%************************************************************************************************************************
%%%%

\subsubsection{Quantum eigenstates and multifractality}

We now turn to the eigenstates. Typical Husimi representations of long-lived eigenstates are shown in \fig{husimi_xop_random}. The spreading of  the distribution in $x$ over the complement of the opening is a consequence of the instantaneous spreading of wavepackets over the size of the system. This spreading washes out the classical structures of the dynamical map as they are now scattered in the $x$ direction. The corresponding $t_E=1$ is minimal, and therefore these systems represent an extreme case of the classical description. 
  %The classical description of Eqs.~\eqref{keating_1} and \eqref{keating_2} in this case with $t_E=1$ leads to the following picture. \mygreen{COMPLETE}
The wavefunctions are strongly localized in $p$ at large $\lam$, and as $\lam$ decreases they become increasingly spread over the whole interval. 

%************************************************************************************************************************
%\subsubsection{Multifractal analysis}

The results of the multifractal analysis are gathered in \fig{mf_xop_random}. In the case of the small opening $\ell=2^{-6}$ ({Figs.~\ref{mf_xop_random}(a-c)}) we see from the behavior of $\widetilde{D}_{2}$ that there is a transition, as $|\lambda_j|$ decreases (top to bottom row), from states that are very localized and with a multifractality present at small scales towards more extended states with multifractal structure at intermediate scales. This general picture is reasonably close to the results for the dynamical map with the same opening, in accordance with the fact that $t_E$ is of the same order in both cases ($t_E=1$ and $t_E=2$, respectively).

For the larger opening $\ell=2^{-2}$ ({Figs.~\ref{mf_xop_random}(h-j)}), while the localization of states for large $\lam$ is more drastic ({Figs.~\ref{mf_xop_random}(h) and \ref{mf_xop_random}(i)}), the states at smaller values of $\lam$ ({Figs.~\ref{mf_xop_random}(j)}) present a clear plateau for $\tilde{D}_2$ over most of the scales, at a value that is close to the value for the closed system. Contrary to the situation for the $\ell=2^{-6}$ case, here the presence of random phases makes the system much more multifractal compared to the dynamical map. This is consistent with the fact that $t_E$ for $\ell=2^{-2}$ is much longer for the dynamical system than for the random one. We see thus that the presence of random phases completely changes this time scale, and is tantamount to the system not being sensitive to any intricate classical structure. 

The results presented show that those properties of the dynamical map that depend on the variations of $t_E$ with the size of the opening, disappear for the random map. Instead, we see that a clear multifractality is visible for both opening sizes, with different behavior with respect to $\lam$. Since random phases can represent the average of quasimomenta present in experimental implementations, this can be seen as a positive sign for the observation of multifractality in realistic open maps.

%As discussed in the previous section, the notion of quantum multifractality is, by definition, directly tied to the distribution of states in a given representation. We are then naturally led to wonder about the interplay between the structures induced by opening the map and those which define any eventual multifractality of the original closed-system eigenstates. As we will see, multifractality in the semiclassical limit will only be possible when the sets $K_{\pm}$ are connected enough in the representation at hand, something which, in our case, happens only when opening along $p$.

%************************************************************************************************************************
%************************************************************************************************************************
\subsection{Opening in momentum}\label{sec:pop}
%************************************************************************************************************************
%************************************************************************************************************************
We now consider the case where the opening is in $p$. According to Eqs.~\eqref{oqmap}--\eqref{proj}, the open quantum map is defined as
\begin{equation}
\widetilde{U} = U\left(1 -\sum_{p=0}^{\lfloor N\ell\rfloor-1} \ket{p}\bra{p}\right)\, ,
\end{equation}
which is realized in momentum representation by simply setting the first $\lfloor N\ell\rfloor$ rows of $U_{pp'}$ to 0.

%A primary question is whether for small enough opening size the resulting system can be understood as a true perturbation of the original one. If it can be established that certain open-system eigenstates are perturbatively away from some closed-system ones then it will be fair to expect that those perturbed states are also multifractal, although perhaps more weakly so than their unperturbed counterparts.

 In this section, we consider, as in the previous studies, the left eigenstates of the open map, and we work exclusively with the random-phase version of \eqref{qmap}. {As mentioned above, the dynamics in $p$ is unaffected by the introduction of random phases, and therefore, when opening in $p$, this is solely for the sake of improving our statistics, and the results to be discussed follow closely those of the dynamical map opened in $p$ with only minor differences regarding the exact values of the $D_q$.}

%************************************************************************************
\subsubsection{Classical properties}\label{sec:mom_classical}
%************************************************************************************

\begin{figure}[!h] %fig9
\includegraphics[scale=1]{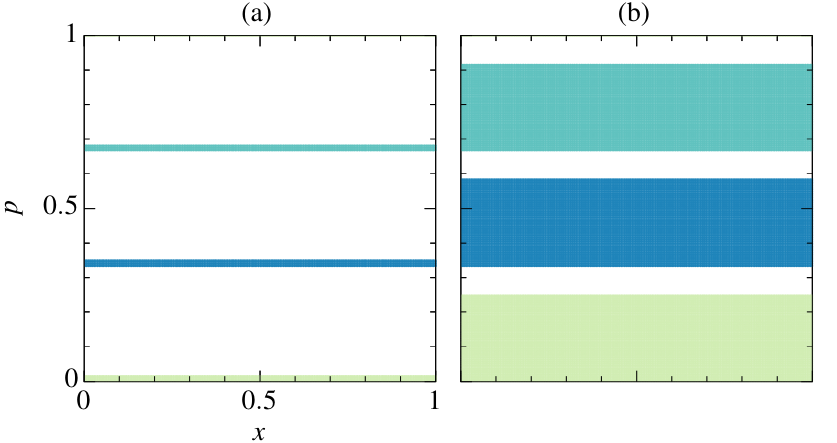}
\caption{(Color online) Regions $R_-^{m}$ ($m=1,2,3$ as blue, blue-green, green, resp.) for the classical map with $\gamma=1/3$, opened in $p$ with \mbox{$\ell=2^{-6}$} (a) and $2^{-2}$ (b).  As for the structures corresponding to forward evolution, we simply have  $\Omega=R_+^0=R_-^3$, $R_+^1=R_-^2$, $R_+^2=R_-^1$ and $K_+=K_-=K_0$.
}
\label{classical_pop}
\end{figure}

For rational $\gamma=1/b$, the $1/b$ periodicity of \eqref{cmap} in $p$ implies that opening it along this direction will yield a relatively simple structure. The opening $\Omega$ and its classical preimages are depicted in \fig{classical_pop}. Points either escape in the first or second iteration or remain in the system forever. The unstable zones (shaded regions) correpond to $\Omega=R_+^0=R_-^3$, $R_+^1=R_-^2$, $R_+^2=R_-^1$. Noting that $R_+^m=R_-^{m+1}=\O$ for $m\geq 3$ and that $t_E$ is infinite for our system, this gives $K^{t_E}_{+}=K^{t_E}_{-}=K_0$, which is the stable zone (white regions).

\begin{figure}[!h] %%fig10
\includegraphics[scale=1]{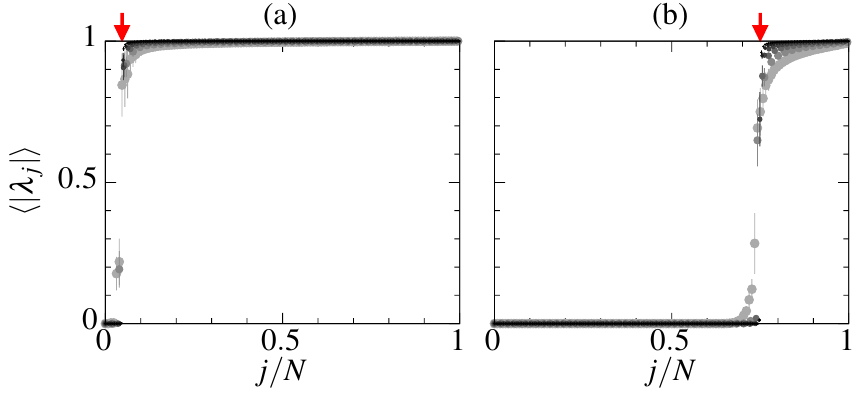}
\caption{(Color online) Average spectral norms $\langle |\lambda_j| \rangle$ of the random map for $\gamma=1/3$, opened in ${p}$ with $\ell=2^{-6}$ (a), $2^{-2}$ (b) and for various system sizes $N=2^{7}$ (lightest gray)$,...,2^{12}$ (darkest gray). For each $N$ we diagonalize $M=2^{17}/N$ random-phase realizations. The index $j\in\{1,...,N\}$ is defined by the ordering $|\lambda_j| \leq |\lambda_{j+1}|$, and the average $\langle |\lambda_j| \rangle$ is taken over the $M$ realizations at fixed $j$. The red arrow at the top of each plot marks the theoretical prediction $\ell b$ for the (rescaled) size of the short-lived sector.}
\label{spectral_norms_pop}
\end{figure}

\begin{figure*}[!t]  %%fig11
\includegraphics[scale=1]{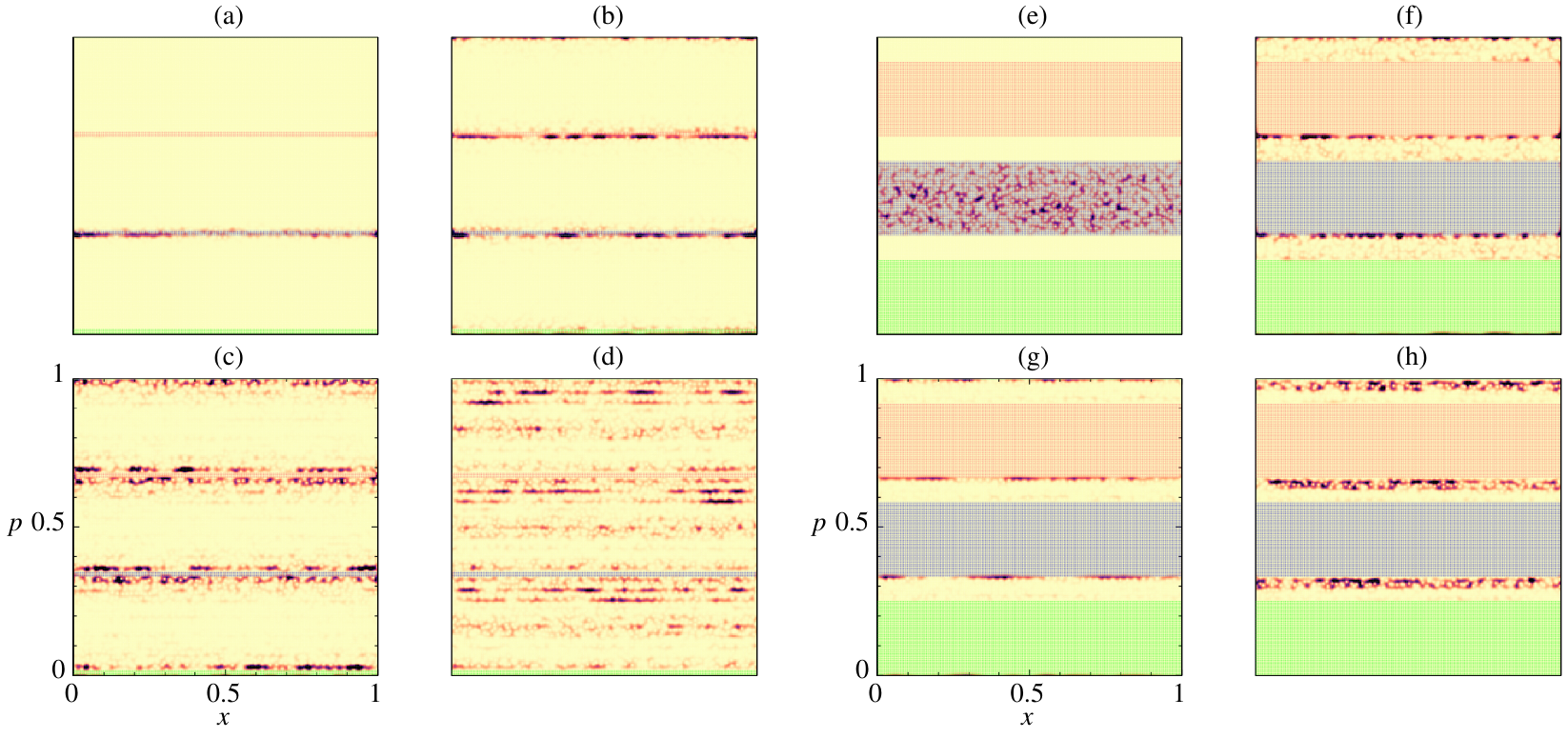}
\caption{(Color online) Husimi representation of four eigenstates of a given realization of the random map opened in $p$ for $\ell=2^{-6}$ (a-d) and $\ell=2^{-2}$ (e-h). Superimposed on each panel are the classical regions $R_-^1$, $R_-^2$ and $R_-^3$ (blue, red, green; dark to light, resp.). The eigenvalue norm $|\lambda|$ associated to each eigenstate is (a) 0.19009, (b) 0.95790, (c) 0.99527, (d) 0.99952, (e) 0.01785, (f) 0.95803, (g) 0.99522, (h) 0.99958.
}
\label{husimi_pop_random}
\end{figure*}
%************************************************************************************************************************
\subsubsection{Spectral properties}\label{sec:mom_classical}
%************************************************************************************************************************

This partioning of phase space, in turn, manifests itself in the spectrum of $\widetilde{U}$, as can be observed in \fig{spectral_norms_pop}, where we display the average spectral norms $\langle |\lambda_j| \rangle$. Again, resonances can be classified as being either short-lived ($\langle |\lambda_j| \rangle\simeq 0$) or long-lived ($\langle |\lambda_j| \rangle\simeq 1$). Following again the analysis of \cite{schomerus}, one can estimate the number of short-lived quantum resonances by the (linear) size of the classical unstable region  (union of shaded regions in \fig{classical_pop}). Here, it is given by $N\ell b$ (where $\gamma=a/b$). This is in agreement with the form of the spectrum (the arrow at the top of the plots in \fig{spectral_norms_pop} corresponds to the rescaled value $\ell b$), and we have checked as well that this relation is satisfied for various other $\ell$ and $\gamma$, with improved agreement as $N\rightarrow\infty$. 
The FWL is trivially verified as the spectral density converges to a step function and  $K_0$ is two-dimensional.
%

%************************************************************************************************************************
%\subsubsection{Quantum states}\label{sec:mom_ipr}
%************************************************************************************************************************
%\mygreen{WHAT DO WE PUT HERE?}
%************************************************************************************************************************
%\subsubsection{IPR: perturbative eigenstates}\label{sec:mom_ipr}
%************************************************************************************************************************
%%
\subsubsection{Quantum states: semiclassical picture and multifractality}\label{sec:mom_ipr}

Again, eigenvectors of the quantum map can be described by their Husimi function, of which four examples are shown {for $\ell=2^{-6}$ in Figs.~\ref{husimi_pop_random}(a-d) and $\ell=2^{-2}$ in Figs.~\ref{husimi_pop_random}(e-h)}. For the largest $\lam$, states are entirely in $K^{t_E}_{0}$. Since the set $\rrm\bigcap K_+^{t_E}$ is empty, as $\lam$ becomes smaller the states tend to settle at the frontier between $\rrm$ and $K_+^{t_E}$. Note that this is not inconsistent with the semiclassical theory insofar as borders are implicitly excluded from our analysis (see Appendix \ref{relativew}). Finally, for very small $\lam$, states spread over $R_-^1$ (seen clearly in {\fig{husimi_pop_random}(e)}). Once more, even though $\rrm\bigcap K_+^{t_E}$ is empty, this does not contradict the theory, since Eq.~\eqref{keating_1} does not apply for small enough $\lam$ (see Eq.~\eqref{a1_minus} and the remark below it).

\begin{figure*}[!t]  %%fig12
\includegraphics[scale=1]{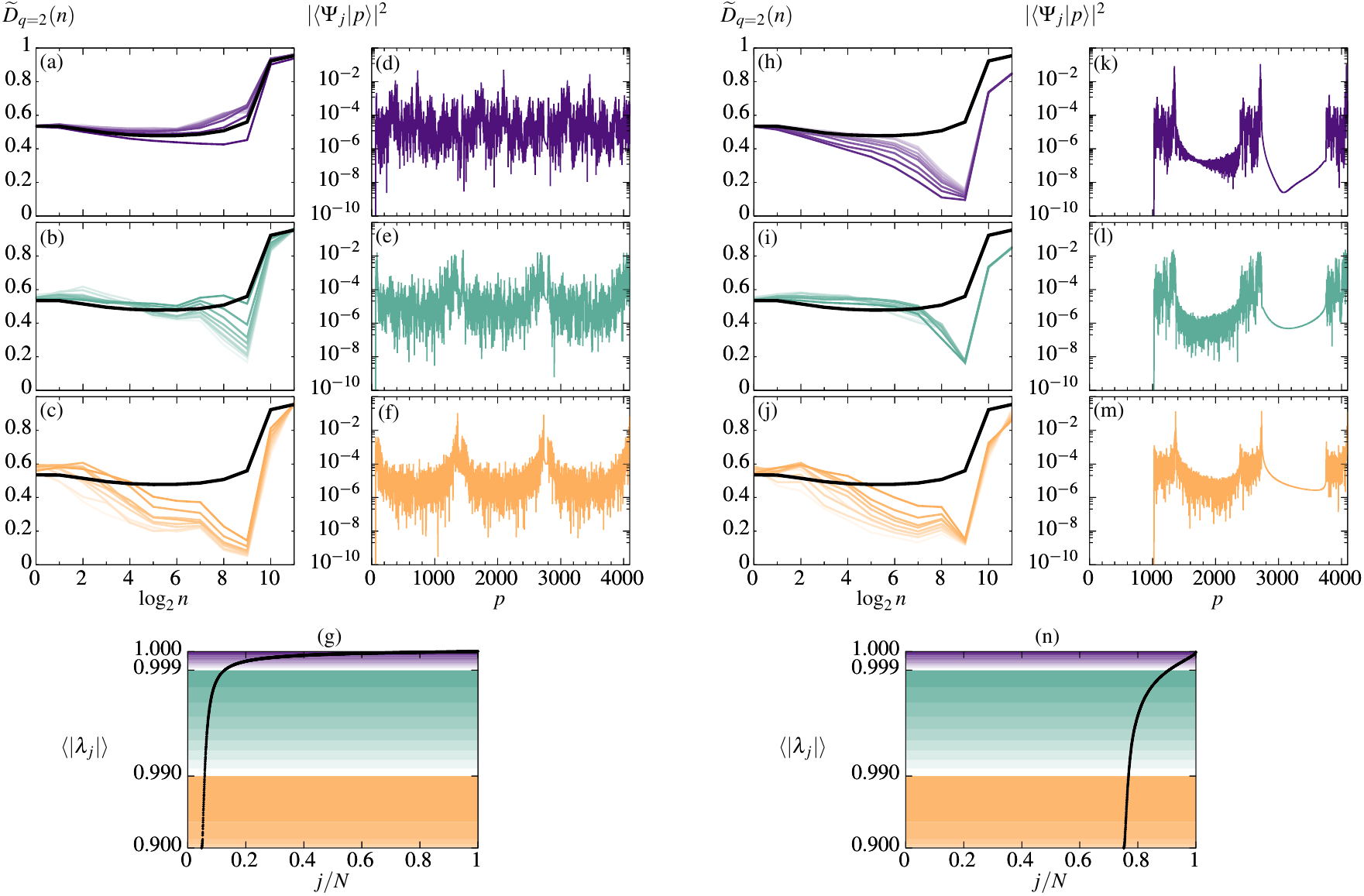}
\caption{(Color online) Multifractal analysis for the random map opened in $p$ with $\ell=2^{-6}$ (a-g) and $\ell=2^{-2}$ (h-n). Top three rows in each case ($\langle|\lambda_j|\rangle$ decreases from top to bottom): box-size $n$ behavior of $\widetilde{D}_2(n)$ for the left eigenstates of the open map $\ket{\Pl_j}$ (thin colored lines) and those of the closed map $\ket{\Pz_k}$ (thick black line); for $\ket{\Pz_k}$ the average is over the $M\times N$ available states, whereas for $\ket{\Pl_j}$ we average over all those states (out of the $M\times N$) falling within a small norm range indicated by the color and shade of the curve in panels (g) and (n) displaying the spectral norms $\langle|\lambda_j|\rangle$ for each case. To the right of each $\widetilde{D}_2(n)$ plot, we display $|\bra{\Pl_j}p \rangle|^2$ for an exemplifying state in the corresponding $\langle|\lambda_j|\rangle$ window. Results correspond to $M=32$ random realizations with $N=2^{12}$ and $\gamma=1/3$.}
\label{mf_pop_random}
\end{figure*}

Our results for the multifractal behavior are summarized in \fig{mf_pop_random} for $\ell=2^{-6}$ and $\ell=2^{-2}$. Fig.~\ref{husimi_pop_random} shows that for $\lam \simeq 1$ the wave functions are located in the set $K^{t_E}_{+}=K_0$, but not in an ergodic way. {Instead, these states show multifractal properties, at least at small scales, as can be seen in {Figs.~\ref{mf_pop_random}(a-c) and \ref{mf_pop_random}(h-j)}. For small opening, the multifractality curves for the largest $\lam$, shown in Fig.~\ref{mf_pop_random}(a), are very close to those of the closed system. We have verified that such states have very strong overlap with certain eigenvectors of the original closed system; they correspond to states of the closed system already localized in the nonescaping regions. For smaller $\lam$, Figs.~\ref{mf_pop_random}(b) and \ref{mf_pop_random}(i) show that multifractality of the open system persists for a range of scales comparable to that of the closed system multifractality, regardless of the size of the opening. Finally, as $\lam$ decreases even further (but remains close to one) the states are at the frontier between $\rrm$ and $K_+^{t_E}$ ({Figs.~\ref{husimi_pop_random}(b) and \ref{husimi_pop_random}(f)}), with a width $\sim \sqrt{\hbar} \sim 1/\sqrt{N}$ in $p$ (i.e. the order of the size of a coherent state), which corresponds to $\log_2 n \sim \log_2 \sqrt{N}= 6$ in {Figs.~\ref{mf_pop_random}(c) and \ref{mf_pop_random}(j)}, where it can be seen that there is clear multifractal behavior below this scale.}

The step-like spectral density makes the intermediate values of $\lam$ not relevant. As for the very small values of $\lam$, they correspond to states concentrated on $R_{-}^1$ (see {Figs.~\ref{mf_pop_random}(a) and \ref{mf_pop_random}(h)}). Our calculations reveal that they are actually close to ergodic on this set and that they have no multifractal properties (data not shown). Note that this resembles the situation encountered when the dynamical map was opened in $x$, where states with small $\lam$ were found to be ergodic within their associated classical structure. This may be a general effect due to the vicinity of the huge degenerate subspace which contaminates this type of state.

%************************************************************************************************************************
%************************************************************************************************************************
\section{Conclusion}
%************************************************************************************************************************
%************************************************************************************************************************
\label{secconc}
We have explored the properties of open quantum maps whose closed counterpart has multifractal properties. We focused on a specific pseudointegrable system, where multifractality is visible in momentum representation. The semiclassical description of resonance eigenstates has been shown to follow the general theory of \cite{Nonnenmacher2005,keating,Nonnenmacher_2007,dima2008,Kopp2010,altmann2013,Korber2013,altmann2015,ketzmerick}, albeit here suitably adapted for the treatment of systems in which one must work at finite $\hbar$. The theory links the phase space distribution of resonance eigenstates to a hierarchical structure arising from the classical partitioning of phase space determined by the escape dynamics: states concentrate on $R_-^{m}\bigcap K_+^{t_E}$, with an increasing visiblity for larger $m$ when the eigenvalue norm $\lam$ increases. For chaotic systems, the quantum states were predicted to be ergodic on average at each hierarchical level. { Our results are consistent with the conjecture that for pseudointegrable systems, individual quantum states are multifractal at each hierarchical level. We have confirmed that multifractality manifests itself for left eigenstates with large and intermediate values of $\lam$ whenever the classical structure has enough space for it to be visible.} For very small $\lam$, the semiclassical behavior of the states is more involved, with states entirely on $R_-^{1}$ or on $R_-^{1}\bigcap K_+^{t_E}$ depending on the opening, and no visible multifractality. In this extreme regime, it must be kept in mind that not only does the general semiclassical theory become inapplicable, but also that such short-lived eigenstates are far from orthogonal and are very close to the huge degenerate space at $\lam = 0$. 

 Instead of focusing on the semiclassical limit of very small $\hbar$, we have been led to adapt the formalism to the case in which $\hbar$ is not asymptotically small compared to the classical structures. In chaotic systems a spectral gap is present that effectively limits the number of structures that are relevant to describe a quantum state. In our case, the absence of a spectral gap means that states with $\lam$ close to unity can have important probability on very small classical structures, which can be comparable to $\hbar$ for arbitrarily small values of the Planck constant. We think that this adapted formalism should be useful for other systems with no spectral gaps or more generally when finite $\hbar$ effects are important.
%{It maybe that the general theory does not apply perfectly in this extreme regime, with very short-lived states which are far from orthogonal and very close to the huge degenerate space at $\lam = 0$.} 

In general, our study shows that the observation of multifractality will depend on $\lam$, which should not be too small, and the Ehrenfest time $t_E$, which should not be too large in order to avoid the presence of too intricate a hierarchical structure in phase space becoming relevant for the quantum system. As $t_E$ is controlled by both the opening size and $\hbar$, these two parameters should be chosen carefully to obtain resonance eigenstates with multifractal properties. Note that for phase-space structures that are of relatively small sizes and compatible with the basis where multifractality manifests itself, it can be seen only at small scales up to a scale given by the classical structure.  Another interesting point is that as there is no gap in the spectrum (contrary to the case of chaotic systems), one can make the hierarchical structure less visible by increasing $\lam$ close to 1.
Finally, although we have confined our study to left eigenstates, we expect our conclusions to hold for right eigenstates as well. As far as multifractality is concerned, we have indeed verified that left and right eigenstates corresponding to a given $\lambda_j$ have very similar behavior.

Our results extend and validate the semiclassical theory of open quantum systems in the case of pseudointegrable systems, and they suggest that it is possible to observe multifractal behavior in open quantum systems, e.g., in scattering experiments, for different kinds of opening.

%************************************************************************************************************************
%\section*{Acknowledgements}
%************************************************************************************************************************
\begin{acknowledgments}
We thank R.~Ketzmerick and A.~B\"acker for discussions. This publication is funded in part by a QuantEmX grant from ICAM and the Gordon and Betty Moore Foundation through Grant GBMF5305 to Olivier Giraud. Agust\'{\i}n Bilen thanks the GDR CNRS 3274 DynQua for financial support, and the French Embassy in Argentina together with the Ministerio de Educaci\'{o}n (Argentina) for funding through the scholarship program ``Becas Saint-Exup\'{e}ry 2018''. We thank CalMiP for access to its supercomputer. This study has been supported through the EUR grant NanoX n° ANR-17-EURE-0009 in the framework of the ``Programme des Investissements d'Avenir'', by  the  ANR  grant 
COCOA  (Grant No. ANR-17-CE30-0024-01), by the CONICET (Grant No. PIP 11220150100493CO), by ANCyPT (Grant No. PICT-2016-1056)
 and by the French-Argentinian LIA LICOQ.
\end{acknowledgments}

\appendix

{

\section{Coherent States on $\mathbb{T}^2$}
\label{appcohstates}
The coherent states $\ket{z}$ used to define the Husimi function on the torus $\mathbb{T}^2$ are coherent states of the harmonic oscillator that are properly periodized as befits the torus structure of the present context \cite{bouzouina1996}. A coherent state of the harmonic oscillator centered at $(X,P)$ has the following expression in $x$ and $p$ representation respectively:
\begin{align}
\label{cohstates1}
\psi_{(X,P)}(x) &= 
\left(\frac{1}{\pi \hbar}\right)^{\frac{1}{4}} 
e^{-\frac{i}{\hbar}\frac{PX}{2}}
e^{\frac{i}{\hbar}{Px}}
e^{-\frac{1}{2\hbar}(x-X)^2}\\
\label{cohstates2}
\hat{\psi}_{(X,P)}(p) &= 
\left(\frac{1}{\pi \hbar}\right)^{\frac{1}{4}} 
e^{\frac{i}{\hbar}\frac{PX}{2}}
e^{-\frac{i}{\hbar}{Xp}}
e^{-\frac{1}{2\hbar}(p-P)^2}\,.
\end{align}
In the position and momentum eigenbases $\{\ket{x_j}\}_{j=0}^{N-1}$ and $\{\ket{p_j}\}_{j=0}^{N-1}$ with $x_j,p_j\in \{0,1/N,\ldots,(N-1)/N\}$ and \mbox{$\hbar_{\rm eff}=\hbar =1/(2\pi N)$}, a coherent state in $\mathbb{T}^2$ is given by
\begin{equation}
|X,P\rangle =
\begin{cases}
\sum_{j=0}^{N-1} c_j(X,P)|x_j\rangle
\\
\\
\sum_{j=0}^{N-1} d_j(X,P)|p_j\rangle
\\
\end{cases}
\end{equation}
with coefficients given by
\begin{align}
c_j(X,P) &=\sqrt{\frac{1}{N}}\sum_{m\in \mathbb{Z}} \psi_{(X,P)}(x_j-m)
 \label{cj}
\\
d_j(X,P) &= \sqrt{\frac{1}{N}}\sum_{m\in \mathbb{Z}}\hat{\psi}_{(X,P)}(p_j-m)\,.
 \label{dj}
\end{align}
Equation \eqref{cj} corresponds to Eq.~(16.26) of \cite{gazeau} (with \mbox{$\kappa_{1,2}=0$}, $Z=i$, $a,b=1$ and after correction of a few misprints).

In terms of the elliptic theta function of the third kind \cite{handbook}, $\vartheta_3$, defined as
\begin{equation}
\vartheta_3(x;\tau) = \sum_{n\in\mathbb{Z}} \tau^{n^2}e^{2inx}\, \qquad (|\tau|<1),
\end{equation}
we can rewrite the coefficients as 
\begin{equation}
c_j(X,P) = 
\left(\frac{2}{N}\right)^{\frac{1}{4}} 
e^{-i \,2\pi N
	P
	\left(
	\frac{X}{2}
	-\frac{j}{N}
	\right)
	}
e^{-\pi N
	\left(
	\frac{j}{N}
	-X
	\right)^2
	}
\vartheta_3(r_j;\tau)
\end{equation}
with $r_j= -\pi N {P}-\pi N {i}(j/N-X)$ and $\tau=\exp(-\pi N)$, and
\begin{equation}
d_j(X,P)= 
\left(\frac{2}{N}\right)^{\frac{1}{4}} 
e^{i\,2\pi N
	X
	\left(
	\frac{P}{2}
	-\frac{j}{N}
	\right)
	}
e^{-\pi N
	\left(
	\frac{j}{N}
	-P
	\right)^2
	}
\vartheta_3(s_j;\tau)
\end{equation}
with $s_j=\pi N X-\pi N i(p_j-P)$. The coherent states defined above satisfy the following resolution of the identity 
\begin{equation}
\int_{\mathbb{T}^2} \frac{dXdP}{2\pi\hbar} |X,P \rangle \langle X,P | = \mathbb{1}.
\end{equation}
In our computations, coherent states are centered at points $(X_i,P_j)\in \Gamma$, where $\Gamma$ is a lattice covering the torus $\mathbb{T}^2$, consisting of $\lfloor \sqrt{2/\hbar}+\frac12\rfloor^2$ square cells of width $\sqrt{\hbar/2}$ whose position is fixed by the first cell centered at $(\sqrt{\hbar/8}, \sqrt{\hbar/8})$. In $\Gamma$ the resolution becomes
\begin{equation}
\frac{1}{4\pi}\sum_{(X_i,P_j)\in \Gamma}  |X_i,P_j \rangle \langle X_i,P_j | = \widetilde{\mathbb{1}} \, ,
\end{equation}
where $\widetilde{\mathbb{1}}$ satisfies $ \widetilde{\mathbb{1}}\ket{\phi}=\ket{\phi}$ (up to corrections of order $\hbar$) for any $\ket{\phi}$.
A projector $\Pi_R$ onto a region $R\in \mathbb{T}^2$ is defined as \cite{saracenovoros,saracenovall} 
\begin{equation}
\label{defproj}
\Pi_R = \frac{1}{4\pi}\sum_{(X_i, P_j)\in \Gamma} \chi_R(X_i,P_j)|X_i,P_j \rangle \langle X_i,P_j |,
\end{equation}
where $\chi_R$ is the characteristic function of $R$:
\begin{equation}
\chi_R(X,P) =
\begin{cases}
1 \,,\,(X,P)\in R \\
0 \,,\,(X,P)\notin R\,.
\end{cases}
\end{equation}
%

%{Finally, we define the \emph{broadened} version $R(\hbar)$ of a region $R$ as
%
%\begin{equation}
%\label{defproj}
%R(\hbar) = \{z\in \mathbb{T}^2: d(z,z')\leq \sqrt{\hbar/2} \text{ for all } z'\in R\},
%\end{equation}
%
%with $d$ the Euclidean distance. We see that, in terms of our lattice, this broadening amounts to an `outwards' thickening of $R$ by one minimal cell. The corresponding projector is then simply
%
%\begin{equation}
%\label{defproj}
%\Pi_{R(\hbar)} = \frac{1}{4\pi}\sum_{(X_i, P_j)\in \Gamma} \chi_{ R(\hbar)}(X_i,P_j)|X_i,P_j \rangle \langle X_i,P_j |.
%\end{equation}
%}

\section{Relative weights on classical sets}
\label{relativew}

We now derive in detail the results quoted in the introduction. We start from the exact relation
\begin{equation}\label{0}
|\lambda_n|^{2m} |\langle \Psi_n^{-} | z \rangle|^2 = 
|\bra{\lev} \widetilde{U}^m |z  \rangle |^2,
\end{equation}
where $|z\rangle$ is a coherent state centered at $z=(X,P)$. We have
\begin{equation}\label{a_evo_minus}
\widetilde{U}^m \ket{z} = \underbrace{ U (\proj)\cdots  U (\proj)}_{m \text{ times}} \ket{z}\,,
\end{equation}
where $\Pi_0=\Pi$ is the projector on the opening $\Omega$.
Points $(X,P)$ starting in the opening will be eliminated by the first projection $(\proj)$, and in the regime where $m<t_E$ only points which have not escaped after $m$ iterations will survive \eqref{a_evo_minus}. Using \eqref{0} we thus obtain that if $|\lambda_n|>0$ then
\begin{equation}\label{a1_minus}
|\bra{\lev} z\rangle|^2 \text{ concentrates on } z\in K_+^{t_E}\equiv \Bigg( \bigcup_{0\leq m < t_E} \rrp \Bigg)^c \, .
\end{equation}
{However, as seen from} \eqref{0} {, for $|\lambda_j|\approx 0$ one may expect $|\bra{\lev} z\rangle|^2$ to have non-negligible support in regions outside of $K_+^{t_E}$.}

Let $A=U\Pi_0 U^{\dagger}$. From the definition $R_-^1=\mathbf{M}\hole$ we have $A\approx \Pi_-^1$ as long as quantum dynamics follows classical dynamics (and provided $t_E>1$). Using the fact that $A= 1-\widetilde{U}\widetilde{U}^{\dagger}$ we have $\langle \Psi_n^{-} | A | \Psi_n^{-}\rangle=1- |\lambda_n|^2$, and thus
\begin{equation}\label{ap_2}
\bra{\lev} \Pi_-^1 \ket{\lev} \approx 1- |\lambda_n|^2   \,.
\end{equation}
In order to proceed further, we use the classical recursive relation for the $R_-^{m}$, which  reads $R_-^{m+1}=\mathbf{M}(R_-^{m}\backslash\hole)$. Together with the definition of $\widetilde{U}$ it implies that for $1\leq m<t_E$ we have $\widetilde{U}^{m-1} \Pi_-^1 (\widetilde{U}^{\dagger})^{m-1}\approx\Pi^{m}_-$, with $\Pi^{m}_-$ being the projector onto $R_-^{m}$. Inserting this into the identity
\begin{equation}\label{1}
|\lambda_n|^{2(m-1)} |\langle \Psi_n^{-} | A | \Psi_n^{-} \rangle| = 
|\bra{\lev} \widetilde{U}^{m-1} A (\widetilde{U}^{m-1})^{\dagger} \ket{\lev} |,
\end{equation}
we get
\begin{equation}\label{s2_minus}
\bra{\lev} \Pi_{m}^{-} \ket{\lev} \approx |\lambda_n|^{2(m-1)}(1-|\lambda_n|^2) \qquad \forall m: 1\leq m < t_E \,.
\end{equation}

}
%%%%%%%%%%%%%%%%%%%%%%%%


%%%%%%%%%%%%%%%%%%%%%%%%%%%%%%

\begin{thebibliography}{74}
\bibitem{footnote} Consider, for simplicity, $R_-^{1}$. The width of its stripes can be obtained by first propagating the left and right edges of the opening, given by $x=0$ (mod 1) and $x=\ell$ (mod 1), resp. The first forward images of these under the map \eqref{cmap} are given by the parametric equations $p(x)=x/2$ (mod 1) and $p(x)=(x-\ell)/2$ (mod 1), and the distance $\delta$ between them gives the sought width of the stripes: $\delta=\ell/\sqrt{5}$. On the other hand, with $N=2^{12}$ we have $\sqrt{\hbar/2}=1/\sqrt{2^{14}\pi}$ so that we may express the width $\delta$ in units of a Planck cell: $\frac{\delta}{\sqrt{\hbar/2}}=2^{7}\ell\sqrt{\pi/5}$. For $\ell=2^{-6}$ this gives $\frac{\delta}{\sqrt{\hbar/2}}\approx 1.59$, i.e. the smallest characteristic size of $R_-^{1}$ is of the order of a Planck cell. For $R_-^m$, this quantity decreases with $m$.%
\end{thebibliography}
\end{document}